\documentclass[fleqn,12pt]{article}
\usepackage{amsfonts,epsfig,latexsym}
\usepackage{amssymb}
\usepackage{graphicx}
\usepackage{amsmath}

\newcommand{\ft}[2]{{\textstyle {\frac{#1}{#2}} }}

\newcommand{\be}{\begin{equation}}
\newcommand{\ee}{\end{equation}}
\newcommand{\ben}{\begin{displaymath}}
\newcommand{\een}{\end{displaymath}}
\newcommand{\bea}{\begin{eqnarray}}
\newcommand{\eea}{\end{eqnarray}}
\newcommand{\nn}{\nonumber}

\newcommand{\bean}{\begin{eqnarray*}}
\newcommand{\eean}{\end{eqnarray*}}
\newcommand{\one}{{\rm 1\kern -.9mm l}}

\makeatletter \@addtoreset{equation}{section} \makeatother

\begin{document}

\begin{titlepage}

    \thispagestyle{empty}
    \begin{flushright}
        \hfill{CERN-PH-TH/2008/218}\\
        \hfill{UCB-PTH-08/74}\\
        \hfill{SU-ITP-08/31}\\
    \end{flushright}

    \begin{center}
        { \LARGE\bf Intersecting Attractors}\vspace{15pt}

         {\large{\bf Sergio Ferrara$^{\diamondsuit\sharp\clubsuit}$,
         Alessio Marrani$^{\heartsuit\clubsuit}$,\\[1ex]
         Jose F.\ Morales$^{\spadesuit}$, and\
         Henning Samtleben$^{\flat}$}}

        \vspace{10pt}

{\small
        {$\diamondsuit$ \it Theory division, CERN, Geneva, Switzerland \\[-.3ex]
   \vspace{1pt}
     {$\sharp$ \it Miller Institute for Basic Research in Science,  Berkeley,  USA}\\[-.3ex]
        \vspace{1pt}
           {$\clubsuit$ \it INFN - LNF,  Frascati,  Italy} \\[-.3ex]
        \vspace{1pt}

  {\small \tt{sergio.ferrara@cern.ch}}}

        \vspace{1pt}

        \vspace{2pt}

        {$\heartsuit$ \it Stanford Institute for Theoretical Physics, Stanford, USA\\[-.3ex]
    \vspace{1pt}
        {\small \tt{marrani@lnf.infn.it}}}


        {$\spadesuit$ \it I.N.F.N. Sezione di Roma  ``Tor Vergata'',  Roma, Italy\\[-.3ex]
\vspace{1pt}
        {\small \tt{francisco.morales@roma2.infn.it}}}

        \vspace{3pt}

        {$\flat$ \it Universit\'e de Lyon, Laboratoire de Physique, ENS Lyon, France\\[-.3ex]
     \vspace{1pt}
          {\small \tt{henning.samtleben@ens-lyon.fr}}}
}

\end{center}


\begin{abstract}

  We apply the entropy formalism to the study of the near-horizon geometry of extremal black $p$-brane  intersections in $D>5$ dimensional
supergravities.

The scalar flow towards the horizon is described in terms an
effective potential given by the superposition of the kinetic
energies of all the forms under which the brane is charged. At the
horizon active scalars get fixed to the minima of the effective
potential and the entropy function is given in terms of U-duality
invariants built entirely out of the black $p$-brane charges.

 The resulting entropy function
 reproduces the central charges of the dual  boundary CFT and gives rise to  a Bekenstein-Hawking like area law.

 The results are illustrated in the case of black holes and black
string intersections in $D=6, 7, 8$ supergravities where the
effective potentials, attractor equations,  moduli spaces and
entropy/central charges are worked out in full detail.
\end{abstract}

\end{titlepage}



 \section{Introduction}


 In $D>5$ dimensions, supergravity theories involve a rich variety of tensors fields of various rank (see \textit{e.g.}\ \cite{Salam:1989fm,ADF-U-duality}).
 A single black hole solution is in general charged under different forms and can be thought of as
 the intersection on a timelike direction of extended branes of various types. More generally,
 branes intersecting on a $(p+1)$-dimensional surface lead to a  black $p$-brane intersecting configuration \cite{Lu:1997hb}.
 In complete analogy with what happens in the case of $D=4, 5$ black holes, one can think of the
 $D>5$ solutions as a scalar attractor flow from infinity to a horizon where a subset of the scalars
 becomes fixed to particular values depending exclusively on the black $p$-brane charges.
 The study of such flows requires a generalization of the attractor mechanism \cite{Ferrara:1995ih,Strom,FK1,FK2,FGK} in order
 to account for $p$-brane solutions carrying non-trivial charges under forms of various rank.
In this paper we address the study of these general attractor flows.

We focus on static, asymptotically flat, spherically symmetric, extremal black $p$-brane solutions in supergravities
at the two derivative level
\cite{Gibbons:1993sv}. The analysis combines standard attractor
techniques based on the extremization of
 the black hole central charge~\cite{Ferrara:1995ih,Strom,FK1,FK2,FGK} and the so-called
 ``entropy function formalism"  introduced in~\cite{Sen:2005wa}
  (see \cite{Bellucci:2007ds,Sen:2007qy} for reviews and complete lists of references).
 Like for black holes carrying vector-like charges,  we define the entropy function for black $p$-branes
  as the Legendre transform with respect to the brane
 charges of the supergravity action evaluated at the near-horizon geometry
 (see \cite{Garousi:2007zb} for previous investigations of black rings and non-extremal branes
 using the entropy formalism).
The resulting entropy function  can be written as a sum of a gravitational term and an
effective potential $V_{\rm eff}$ given as a superposition of  the
kinetic energies of the forms under which the brane is charged.
Extremization of this effective potential gives rise to the attractor equations
which determine the values of the scalars at the horizon
as functions of the brane charges. In particular, the entropy
function itself can be expressed in terms of the
U-duality invariants built from these charges and it is proportional to the central charge of the dual
 CFT living on the AdS boundary. The attractor flow can then be thought of as a $c$-flow towards the
 minimum of the {\it supergravity c-function}  \cite{Freedman:1999gp,Goldstein:2005rr}.
 Interestingly, the central charges for extremal black $p$-branes satisfy an area law formula generalizing
 the famous Bekenstein-Hawking result for black holes.

We will illustrate our results in the case of extremal black holes and black strings in $D=6,7,8$ supergravities.
In each case we derive the entropy function $F$ and the near-horizon geometry via extremization of $F$.
At the extremum, the entropy function results into a U-duality invariant combination of the brane charges
reproducing the black hole entropy and the black string central charge, respectively.
Scalars fall into two classes: ``fixed scalars" with strictly positive masses and ``flat scalars"
not fixed by the attractor equations, which span
the moduli space  of the solution.
The moduli spaces will be given by symmetric product spaces that can be
interpreted as the intersection of the charge orbits of the various branes entering in the solution.
In addition  one finds extra ``geometric moduli" (radii and Wilson lines) that are not fixed by the attractors.

The paper is organized as follows.
In Sect.~\ref{Sect2} we derive a Bekenstein-Hawking like area law for central
charges associated to extremal black $p$-branes. In Sect.~\ref{Sect3} the ``entropy function''
formalism is adapted to account for solutions charged  under forms of
different rank. In Sect.~\ref{Sect4} we anticipate and summarize
in a very universal form the results for the set of
theories considered in detail in the
rest of the paper, namely the two non-chiral $\left( 1,1\right) $ and
$\left( 2,2\right) $ supergravities in $D=6$ (Sects.~\ref{Sect5} and
\ref {Sect6}, respectively), and the maximal $D=7$, $8$ supergravities
(Sects.~\ref {Sect7} and \ref{Sect8}, respectively). In Sect.~\ref{Sect9}
the uplift of the previously discussed near-horizon
geometries to $D=11$ $M$-theory is briefly discussed. The concluding
Sect.~\ref{Sect10} contains some final remarks and comments.


 \section{Area Law for Central Charges}\label{Sect2}


 Before specifying to a particular supergravity theory, here we
 derive a universal  Bekenstein-Hawking like formula underlying any gravity flow
 (supersymmetric or not) ending on an AdS point.
Let $AdS_{d}\times \Sigma_m$, with $\Sigma_m$ a product of Einstein
spaces, be the near-horizon geometry of an extremal black $(d-2)$-brane
solution in $D=d+m$ dimensions. After
reduction along $\Sigma_m$ this solution can be thought as the vacuum
of a gauged gravity  theory in $d$ dimensions. To keep the discussion, as general as possible, we analyze the solution
from its $d$-dimensional perspective.
The only fields that can be turned on consistently with
 the $AdS_d$ symmetries are constant scalar fields. Therefore we
can describe the near-horizon dynamics in terms of a gravity theory coupled to scalars $\varphi^i$
with a potential~$V_d$. The potential $V_d$ depends on the details of the higher-dimensional theory.
The ``entropy function" is given by evaluating this action at the $AdS_d$ near horizon geometry (with constant scalars $\varphi^i\approx u^i$)
   \be
   F~=~-\frac1{16 \pi G_{d}} \int d^dx \sqrt{-g} \,(R-V_d)~=~
   \frac{\Omega_{AdS_d}  \,r_{\!\rm AdS}^d }{16 \pi G_{d} } \left\{ \frac{d(d-1)}{r_{\!\rm AdS}^2}+V_d\right\}
   \;,
   \label{FF}
   \ee
  with $ r_{\!\rm AdS}  $ the AdS radius and $\Omega_{AdS_d} $ the  regularized volume of an AdS slice of radius one.
   Following  \cite{Sen:2008vm}  we take for $\Omega_{AdS_d} $  the finite part of the AdS volume integral when the cut off is sent to infinity.
   More precisely we write the AdS metric
   \be
   ds^2= r_{\!\rm AdS}^d ( d\rho^2-\sinh^2\!\rho \, d\tau^2+ \cosh^2\! \rho\, d\Omega_{d-2}^2)
   \;,
   \ee
   with $\tau\in[0,2\pi]$, $0\leq \rho \leq \cosh^{-1} r_0$ and
   $d\Omega_{d-2}$ the volume form of a unitary (d-2)-dimensional sphere. The regularized volume
  $ \Omega_{AdS_d}  $ is then defined as the (absolute value of the)
  finite part of the  volume integral  $\int d^dx\sqrt{-g}$ in the limit $r_0\to \infty$. This results into
    \be
  \Omega_{AdS_d}   = {2\pi\over (d-1)} \Omega_{d-2}
  \;.
  \label{reg_vol}
  \ee
   A different prescription for the volume regularization leads to a redefinition of the entropy function by a
   charge independent irrelevant constant.
The ``entropy" and near-horizon geometry follow from the extremization of
the entropy function $F$ with respect to the fixed scalars $u^i$ and the radius $r_{\!\rm AdS}$
  \bea
  \frac{\partial F}{\partial u^i}~\propto~   \frac{\partial V_d}{\partial u^i}
  &\stackrel!{\equiv}&0
  \;,\nn\\
   \frac{\partial F}{\partial r_{\!\rm AdS} }~\propto~   r^2_{\rm AdS}\, V_d
   +(d-1)\,(d-2)&\stackrel!{\equiv}&0\;.
  \eea
The first equation determines the values of the scalars at the horizon. The second equation determines
the radius of AdS in terms of the value of the potential at the minimum. Notice that solutions exist only if the potential $V_d$
is negative.
Indeed, as we will see in the next section, $V_d$ is always composed from a part
proportional to a positive definite effective potential $V_{\rm eff}$
generated by the higher dimensional  brane charges and a negative
contribution $-R_{\Sigma}$ related to the constant curvature
of the internal space $\Sigma$ (see eq.(\ref{veffv}) below) .
The ``entropy" is given by
evaluating $F$ at the extremum and can be written in the suggestive form
\bea
F&=&
\frac{ \Omega_{d-2}\, \, r_{\!\rm AdS}^{d-2}}{4 \, G_{d} }\ ~=~
\Omega_{d-2}\, r_{\!\rm AdS}^{d-2}\, \, \frac{  A }{4\, G_{D} }
\;,\label{cc}
\eea
 where $A$ denotes the area of $\Sigma_m$,
 $\Omega_{d-2}$ is the volume of the unit (d-2)-sphere,
  and $G_D=A G_d$ the $D$-dimensional Newton constant.
 For black holes ($d=2$), this formula is nothing than the well known
   Bekenstein-Hawking entropy formula $S=\frac{A}{4 G_D}$ and it shows that
 $ F$ can be identified with the black hole entropy.
  For black strings ($d=3$),
 ${3\over \pi}  F=\frac{3 r_{\!\rm AdS}}{2 G_3} $ reproduces the central charge  $c $
  of the two-dimensional CFT  living on the $AdS_3$ boundary \cite{Brown:1986nw}.
  In general, the scaling of (\ref{cc}) with the AdS radius matches that of the
  supergravity {\it c-function} introduced in  \cite{Freedman:1999gp}
 and it suggests that $F$ can be interpreted as the critical value of  the central charge $c$
 reached at the end of the attractor
  flow.

 In the remainder of this paper we will study the flows from the
 $D$-dimensional perspective where
 the black $p$-branes  carry in general charges under forms of various rank.


   \section{The Entropy Function }\label{Sect3}


  The bosonic action of supergravity in $D$-dimensions can be written as
\be
 S_{\rm SUGRA}= \int \left( R *\one-\ft12 \,g_{ij}(\phi) \,d\phi^i\wedge * d\phi^j
  -\ft12 \,N_{\Lambda_n \Sigma_n}(\phi^i)\,
 \,F^{\Lambda_n}_n \wedge * F_n^{\Sigma_n} +{\cal L}_{\rm WZ}\label{action0}
    \right)
    \; ,
\ee
with  $F^{\Lambda_n}_n$,  denoting a set of $n$-form field strengths, $\phi^i$ the scalar fields
living on a manifold with metric $g_{ij}(\phi)$ and  ${\cal L}_{\rm WZ}$
some Wess-Zumino type couplings.
The scalar-dependent positive definite matrix $N_{\Lambda_n \Sigma_n}(\phi^i)$
provides the metric for the kinetic term of the $n$-forms.
The sum over $n$ is understood.  In the following we will omit the subscript $n$  keeping
in mind that both the rank of the forms and the range of the indices $\Lambda$
depends on $n$.
We will work in units where $16 \pi G_D=1$, and restore at the end  the dependence on $G_D$.
For simplicity we will restrict ourselves here to solutions
with trivial  Wess-Zumino contributions and this term will be discarded in the following.

 We look for extremal black $p$-brane  intersections  with near-horizon
geometry of topology $M_D=AdS_{p+2}\times S^{m}\times T^{q}$.
Explicitly we look for solutions with near-horizon geometry
\bea
ds^2 &=& r^2_{\!\rm AdS}\, ds^2_{\!\rm AdS_{p+2}}+ r^2_{\!S}\, ds^2_{S^{m}}+ \sum_{k=1}^q r_k^2\,d\theta_k^2\;,\nn\\[.5ex]
F^{\Lambda} &=& p_a^{\Lambda} \alpha^a+  e^{\Lambda r} \beta_{ r}\;, \qquad
\phi^i~=~ u^i
\;,
\label{near}
\eea
with $\vec{r}=(r_{\!\rm AdS},r_{\!S},r_k)$, describing the $AdS$ and sphere radii,
and $u^i$ denoting the fixed values of the scalar fields at the horizon.
$\alpha^a$ and $\beta_{ r}$  denote the   volume forms of the
compact  \{$\Sigma^a$\} and non-compact  \{$\Sigma_{r}$\} cycles, respectively, in  $M_D$.
The forms  are normalized  such as
\bea
\int_{\Sigma^a } \alpha^b=\delta^b_a\;, \qquad
 \int_{\Sigma_{ r} } \beta_{ s}=\delta^r_s\;.
\eea
They define the volume dependent functions $C^{ab},C_{rs}$
\bea
\int_{M_D} \alpha^a \wedge *\alpha^b=C^{ab}
\;,\qquad
\int_{M_D}  \beta_{ r} \wedge *\beta_{ s}=C_{rs}
\;,
\eea
describing the cycle intersections.
In particular, for  the factorized products of AdS space
and spheres we consider here,
these functions are  diagonal matrices with entries
\bea
C^{ab}=\delta^{ab}\,\frac{  v_D}{{\rm vol} (\Sigma^a)^2} \;, \quad\quad
C_{rs}=\delta_{rs}\,      \frac{v_D}{{\rm vol} (\Sigma^r)^2}  \label{cvol}
\;,
\eea
with $v_D$  the volume of $M_D$.
Integrals over AdS  spaces are cut off to a finite volume,
according to the discussion around (\ref{reg_vol}).

The solutions will be labeled by their electric $q_{I r}$ and magnetic charges $p_a^{I}$
defined as
\bea
p_a^{\Lambda} &=&\int_{\Sigma^a } F^{\Lambda}\;, \nn\\
q_{\Lambda r} &=& \int_{*\Sigma^r } N_{\Lambda \Sigma} *F^{\Sigma} =C_{rs} N_{\Lambda \Sigma} e^{\Sigma s}
\;,
 \label{elmag}
\eea
where we denote by  $*\Sigma^r$ the
 complementary  cycle to $\Sigma^r$ in $M_D$.

Let us now consider the ``entropy function"  associated to a black
$p$-brane solution with near-horizon geometry (\ref{near}).
  The entropy  function $F$ is defined as the Legendre transform in the
  electric charges $q_{\Lambda r}$ of $S_{\rm SUGRA}$ evaluated
  at the near-horizon geometry
  \bea
  F &=&  e^{\Lambda r} \, q_{\Lambda r}-S_{\rm SUGRA}\nn\\
  &=&   e^{\Lambda r}\, q_{\Lambda r}-       R\, v_D   +\ft12 \,N_{\Lambda \Sigma}\,
 \,p_a^{\Lambda}  p_b^{\Sigma} C^{ab}-\ft12 \,N_{\Lambda \Sigma}
 \,e^{\Lambda r}  e^{\Sigma s} \,C_{rs}
 \;,
  \eea
     The fixed values  of $\vec{r},u^i,e^{I r}$ at the horizon can be found via
     extremization
      of $F$ with respect to $\vec{r}$, $u^i$, and $e^{I r}$:
\be
 \frac{\partial F}{\partial \vec r} ~=~ \frac{\partial F}{\partial u^i}
 ~=~ \frac{\partial F}{\partial e^{\Lambda r} }~=~0
 \;.
  \ee
From the last equation one finds that
  \be
  q_{\Lambda r}=     \,N_{\Lambda \Sigma}\,
 e^{\Sigma s}\, C_{rs}
 \;,
 \label{defq}
  \ee
in agreement with the definition of electric charges (\ref{elmag}).
Solving this set of equations for $e^{\Lambda r}$ in favor of
$q_{\Lambda r}$ one finds
    \be
 F(Q,\vec{r},u^i)=  - R(\vec{r})\, v_D(\vec{r})
 +\ft12 Q^T \cdot M(\vec{r},u^i)\cdot Q
 \;,
\ee
 with
 \bea
 M(\vec{r},u^i)&=&\left(
\begin{array}{cc}
  N_{\Lambda \Sigma}(u^i) C^{ab} (\vec{r})& 0 \\
  0 & N^{\Lambda \Sigma}(u^i) C^{rs}(\vec{r})
\end{array}
\right)
\;,
 \quad\quad
 Q\left(
\begin{array}{c}
  p^{\Sigma}_a \\
  q_{\Sigma r}
\end{array}
\right)\label{MQ}
\;,
\eea
and $N^{\Lambda \Sigma}$, $C^{rs}$
denoting the inverse of $N_{\Lambda \Sigma}$ and $C_{rs}$ respectively.

It is convenient to introduce the scalar and form intersection ``vielbeine"
${\cal V}_{\Lambda}{}^M, J^{ab}, J^{'rs}$ according to
\bea
N_{\Lambda \Sigma} &=& {\cal V}_{\Lambda}{}^M{\cal V}_{\Sigma}{}^N\,\delta_{MN} \;,\quad\quad
~~
C^{ab} = J^{ac}J^{bc}\;, \quad\quad
~~~C^{rs}= J'{}^{rt} J'{}^{st}
\;.
\label{relations}
\eea
From (\ref{cvol}) one finds
for the factorized products of AdS space and spheres
\bea
J^{ab}=\delta^{ab}\,\frac{v_D ^{1/2}}{{\rm vol} (\Sigma^a)} \;, \quad\quad
J'{}^{rs}=\delta^{rs}\,      \frac{ {\rm vol}(\Sigma^r)}{v_D ^{1/ 2} } \;.  \label{cvol2}
\eea
 The  electric and magnetic central charges can be written in terms of these quantities as
\be
Z_{\rm mag}^{M a}={\cal V}_\Lambda{}^M\, J^{ba}\, p^{\Lambda}_b \;,\quad\quad ~~~~~~~~~~~~
   Z^{r}_{{\rm el},M}=({\cal V}^{-1})_M{}^\Lambda\, \,J'{}^{sr} \,q_{\Lambda s}
   \;.
    \label{zcentral}
\ee
 Combining  (\ref{relations}) and  (\ref{zcentral})
one can rewrite the scalar dependent part of the
entropy function as the effective potential
\be
 V_{\rm eff}~=~
 \ft12 \,Q^T \cdot M(\vec{r},u^i)\cdot Q~=~
 \ft12\,Z_{\rm mag}^{M a}\,Z_{\rm mag}^{M a} +
 \ft12\, Z^{r}_{{\rm el},M}\,  Z^{r}_{{\rm el},M}
 \;.
\ee

For the $n=D/2$-forms in even dimensions the argument is similar,
except for the possibility of an additional topological term
\be
 S_{\rm SUGRA}=  \int
 \Big(R\, *\one-\ft12 \, {\cal I}_{\Lambda\Sigma}(\phi^i)\,
 \,F_n^{\Lambda} \wedge * F_n^{\Sigma}
 -\ft12 \,{\cal R}_{\Lambda\Sigma}(\phi^i)\,
 \,F_n^{\Lambda} \wedge  F_n^{\Sigma}
    \Big)
    \;,
\ee (note that ${\cal R}_{\Lambda\Sigma}=\epsilon{\cal
R}_{\Sigma\Lambda}$, with $\epsilon=(-1)^{[D/2]}$).
 Following the same steps as before  one finds
\bea
 V_{\rm eff}  \ft12\,Q^T \cdot M(\vec{r},u^i)\cdot Q
 \;,
\eea
with
\bea
{M}(\vec{r},u^i) &\equiv&
C^{ab} \left(
\begin{array}{cc}
({\cal I}+\epsilon{\cal R}{\cal I}^{-1}{\cal R})_{\Lambda\Sigma}
&\epsilon( {\cal R}{\cal I}^{-1})_\Lambda{}^\Sigma \\
 ({\cal I}^{-1}{\cal R})^\Lambda{}_\Sigma &
 ({\cal I}^{-1})^{\Lambda\Sigma}
\end{array}
\right)
\;,
\qquad
 Q\left(
\begin{array}{c}
  p^{\Lambda}_a \\
  q_{\Lambda a}
\end{array}
\right)
\;.
\eea
For ${\cal R}=0$ we are back to the diagonal matrix (\ref{MQ}).
In general, thus we obtain for the $D/2$-forms an effective potential
\bea
 V_{\rm eff}&=&
 \ft12\, Q^T \cdot M(\vec{r},u^i)\cdot Q
 ~=~ \ft12\, Z^{Ma} \, Z^{Ma}
 \;,
\eea
with
\bea
Z^{Ma} &=& J^{ba}\,( {\cal V}_{\Lambda}{}^{M}\,p^\Lambda_b
+  {\cal V}^{\Lambda\,M}\,q_{\Lambda b})
\;,
\label{centrald}
\eea
 where ${\cal V}_{I}{}^M=({\cal V}_{\Lambda}{}^M,{\cal V}^{\Lambda\,M})$
 is the coset representative.

Summarizing, in the case of a general supergravity with bosonic action (\ref{action0})
the entropy function is given by
      \be
 F(Q,\vec{r},u^i)=  -R(\vec{r}) v_D(\vec{r}) +V_{\rm eff}(u^i,\vec{r})
 \;,
 \label{FFF}
    \ee
with the
\textit{intersecting-branes effective potential}
 \bea
 V_{\rm eff} &=&
 \ft12\sum_n \, Q_n^T \cdot M_n(\vec{r},u^i)\cdot Q_n \nn\\
 &=& \ft12\, Z^{Ma} \, Z^{Ma}
+
\ft12 \sum_{n\neq D/2}\left( Z_{\rm mag}^{ M_n a_n}\,Z_{\rm mag}^{ M_n a_n}+
 Z^{r_n }_{{\rm el},M_n}\, Z^{r_n }_{{\rm el},M_n}\right)
 \;,
 \label{Veff}
\eea
 where the first contribution in the second line
 comes from the $n=D/2$ forms. Notice that there
 are two types of interference between the potentials coming
 from forms of different rank:
 First, they in general depend on
 a common set of scalar fields and second, they carry a non-trivial
 dependence on the $AdS$ and the sphere radii.
    Besides this important difference the critical points of the effective potential can be
studied with the standard attractor techniques  for vector like charged black holes.

The near-horizon geometry follows from the extremization equations
  \bea
 \nabla V_{\rm eff}\equiv \partial_{u^i}V_{\rm eff} \, du^i \ft12 \sum_n Q_n^T\cdot \nabla M_n(\vec{r},u^i)\cdot Q_n
&\stackrel!{\equiv}&0\;,
\label{attractor}\\
\partial_{\vec r} \left[   -R(\vec{r})\,v_D(\vec{r})  +V_{\rm eff}(u^i,\vec{r})
\right]&\stackrel!{\equiv}&0
\;.
\label{extremize}
\eea
We conclude this section by noticing that after reduction to $AdS_d$, the $D$-dimensional effective
potential $V_{\rm eff}$ combines with the contribution coming from the scalar curvature $R_\Sigma$
of the internal manifold into the $d$-dimensional scalar potential
\be
  V_d =  {1\over v_D} V_{\rm eff} - R_{\Sigma}   \label{veffv}
\ee
   appearing in (\ref{FF}). Notice that the resulting potential is not positive defined and therefore
   an AdS vacuum is supported.


\section{Summary of Results}\label{Sect4}


Before entering into the detailed analysis  of the entropy  function and its minima,
here we summarize our main results
in a universal form independent of the particular dimension $D$ considered.
 We consider extremal black p-brane solutions with $p=0,1$ in $d=6,7,8$ maximal supergravities
 and ${\cal N}=(1,1)$ supergravity in six dimensions . The attractor mechanism for black strings in
  ${\cal N}=(1,0)$ six-dimensional supergravity was studied in \cite{FG2}.

There are three classes of extremal black $p$-brane intersections.
The corresponding near-horizon geometries,
effective potentials $V_{\rm eff}$,
and entropy  functions in each case are given as follows:
\begin{itemize}

\item{$AdS_3\times S^3\times T^n$:
\bea
V_{\rm eff} &=&\frac{v_{\!\rm AdS_3}}{ v_{S^3}} \,| {\bf I}_2|
\;,
\qquad
r_{\!\rm AdS}~=~ r_{S}~=~\frac{| {\bf I}_2|^{1/4}}{2\pi v_{T^n}^{1/4}}
\;,\nn\\[1ex]
F &=& v_D \left( \frac{6}{r_{\!\rm AdS}^2}-\frac{6}{r_{S}^2}\right)+V_{\rm eff}
~=~ | {\bf I}_2|
\;.
 \label{sumads3s3}
\eea
}

\item{$AdS_3\times S^2\times T^n$:
\bea
V_{\rm eff} &=&\ft32 \,\frac{v_{\!\rm AdS_3}}{v_{S^2}} \,v_{T^n}^{1/3} \,| {\bf I}_3|^{2\over 3}
\;,\qquad
 r_{\!\rm AdS}~=~ 2 r_{S}~=~\frac{| {\bf I}_2|^{1/3}}{2\pi v_{T^n}^{1/3}}
 \;,\nn\\[1ex]
F &=& v_D \left( \frac{6}{r_{\!\rm AdS}^2}-\frac{2}{r_{S}^2}\right)+V_{\rm eff}
~=~ | {\bf I}_3|\;.
\nn\\
 \label{sumads3s2}
\eea
}

\item{$AdS_2\times S^3\times T^n$:
\bea
V_{\rm eff} &=&\ft32 \,\frac{v_{\!\rm AdS_2}}{ v_{S^3}  \,v_{T^n}^{1/3}}  \,| {\bf I}_3|^{2\over 3}
\;,\qquad
 r_{\!\rm AdS}~=~ \ft12 r_{S}=\frac{| {\bf I}_3|^{1/6}}{ 2\pi v_{T^n}^{1/3}}\;,
 \nn\\[1ex]
F &=& v_D \left( \frac{2}{r_{\!\rm AdS}^2}-\frac{6}{r_{S}^2}\right)+V_{\rm eff}= | {\bf I}_3|^{1/2}
\;.
\label{sumads2s3}
\eea
}
\end{itemize}
 where
 \bea
  v_D &=&  v_{\!\rm AdS_d} \,  v_{S^m}\,v_{T^n}\nn\\
 v_{\!\rm AdS_d} &=& \Omega_{AdS_d} \, r_{\!\rm AdS}^d   \;,  \qquad
 v_{S^n}~=~r_{S}^n\, \Omega_n \;,\qquad
 v_{T^n}~=~\Omega_1^n\, \prod_{i=1}^n r_i\;, \nn\\
 \Omega_1& =& 2 \pi \;, \quad \Omega_2~=~4 \pi \;,\quad \Omega_3~=~2 \pi^2 \;, \quad
  \Omega_{AdS_2}~=~2 \pi \;,       \quad    \Omega_{AdS_3}~=~2 \pi^2 \;,
 \eea
 are the volumes of the near-horizon AdS/spheres and  ${\bf I}_{2,3}$ are the
 relevant quadratic
 and cubic U-duality invariants built out of the black $p$-brane charges. We stress that
 these invariants involve, in general, charges under forms of various ranks.
 This is also the case for the effective
 potential $V_{\rm eff}$ resulting from the interfering superpositions of the
 various form contributions.

 We also note that in all cases the radii of the circles of the torus $T^n$
 are not fixed by the extremization equations
 but remain as free parameters.

The results (\ref{sumads3s3})--(\ref{sumads2s3}) shows that the
entropy  function $F$  can be related to the black hole entropy and
black string central charges \bea S_{\rm
black-hole}&=&  F~=~| {\bf
I}_3|^{1/2}~=~\frac{v_{S^3} \, v_{ T^n}}{4 G_{D}}
\;,\nn\\[1ex]
c_{\rm black-string}&=&\frac{3}{\pi}\, F~=~\frac{3}{\pi }\, |
{\bf I}_{2,3}|~=~\frac{3 r_{\!\rm AdS}}{2 G_{3}} \;. \eea

In the following we will derive these results
from the corresponding   supergravities
in various space-time dimensions.


\section{${\cal N}=(1,1)$ in $D=6$}\label{Sect5}



\subsection{${\cal N}=(1,1)$, $D=6$ Supersymmetry Algebra}


The half-maximal $\left( 1,1\right) $, $D=6$ Poincar\'{e}
supersymmetry
algebra has Weyl pseudo-Majorana supercharges and $\mathcal{R}$-symmetry $%
SO\left( 4\right) \sim SU\left( 2\right) _{L}\times SU\left(
2\right) _{R}$. Its central extension reads as follows (see
\textit{e.g.}\
\cite{Strathdee:1986jr,VanProeyen:1999ni,Townsend:1995gp})
\begin{eqnarray}
\left\{ \mathcal{Q}_{\gamma }^{A},\mathcal{Q}_{\delta
}^{B}\right\} &=&\gamma _{\gamma \delta }^{\mu }Z_{\mu }^{\left[
AB\right] }+\gamma
_{\gamma \delta }^{\mu \nu \rho }Z_{\mu \nu \rho }^{\left( AB\right) }; \\
\left\{ \mathcal{Q}_{\dot{\gamma}}^{\dot A},\mathcal{Q}_{\dot{\delta}%
}^{\dot B}\right\}  &=&\gamma _{\dot{\gamma}\dot{\delta}\ }^{\mu
}Z_{\mu }^{\left[ \dot A \dot B\right] }+\gamma _{\dot{\gamma}\dot{%
\delta}\ }^{\mu \nu \rho }Z_{\mu \nu \rho }^{\left( \dot A \dot B\right) }; \\
\left\{ \mathcal{Q}_{\gamma
}^{A},\mathcal{Q}_{\dot{\delta}}^{\dot A}\right\}  &=&C_{\gamma \dot{\delta}}Z^{A\dot A}+\gamma _{\gamma \dot{%
\delta}\ }^{\mu \nu }Z_{\mu \nu }^{A\dot A},
\end{eqnarray}
where $A,\dot A=1,2$, so that the (L,R)-chiral supercharges are $%
SU(2) _{\left( L,R\right) }$-doublets.

Notice that, in our analysis of both $\left( 1,1\right) $ and
$\left( 2,2\right) $ $D=6$ supergravities, it holds that $Z_{\mu
\nu \rho }^{\left( AB\right) }=Z_{\mu \nu \rho }^{\left( \dot A \dot B\right) }=0$, because the presence of the
term $Z_{\mu \nu \rho }^{\left( AB\right) }$ is inconsistent with
the bound $p\leqslant D-4$, due to the assumed asymptotical
flatness of the (intersecting) black $p$-brane space-time
background.

Strings can be dyonic, and are associated to the central charges $
Z_{\mu }^{\left[
AB\right] },Z_{\mu }^{\left[ \dot A \dot B\right] }$ in
the $\left( \mathbf{1},\mathbf{1}%
\right)$ of the $\mathcal{R}$-symmetry group. They are embedded in
the $\mathbf{1}_{\pm}$
(here and below the subscripts denote the weight of $SO(1,1)$) of the $U$-duality group $%
SO\left( 1,1\right) \times SO\left( 4,n_{V}\right) $. On the other
hand, black holes and their magnetic duals (black $2$-branes) are
associated to $Z^{A\dot A},Z_{\mu \nu }^{A\dot A}$ in the $\left(
\mathbf{2},\mathbf{2}^{\prime }\right) $ of $SO\left( 4\right) $,
and they are embedded in the $\left(
\mathbf{n}_{V}+\mathbf{4}\right) _{\pm\frac{1}{2}}$ of $SO\left(
1,1\right) \times SO\left( 4,n_{V}\right) $.

In our analysis, the corresponding central charges are denoted
respectively
by $Z_{+}$ and $Z_{-}$ for dyonic strings, and by $Z_{{\rm el},A\dot{A}}$ and $%
Z_{{\rm mag},A\dot{A}}$ for black holes and their magnetic duals.

\subsection{${\cal N}=(1,1)$, $D=6$ Supergravity}

The bosonic field content of half-maximal ${\cal N}=(1,1)$  supergravity in $D=6$
dimensions coupled to  $n_{V}$  {matter} (\textit{%
vector}) multiplets consists of a graviton, $(n_V+4)$ vector fields with field
strengths $F_2^M$, $M=1, \dots, (n_V+4)$,
a three form field strength $H_3$, and $4n_V+1$ scalar fields parametrizing
the scalar manifold
  \begin{equation}
{\cal M}~=~SO\left( 1,1\right) \times \frac{SO\left( 4,n_{V}\right) }{SO\left( 4\right)
\times SO\left( n_{V}\right) }\;,\qquad
{\rm dim}_{\mathbb{R}}\,{\cal M}=4n_{V}+1\;,
\end{equation}
with the dilaton $\phi $ spanning $SO\left( 1,1\right) $, and the $4n_{V}$
real scalars $z^{i}$ ($i=1, \dots, 4n_{V}$) parametrising the quaternionic
manifold $\frac{SO\left( 4,n_{V}\right) }{SO\left( 4\right) \times SO\left(
n_{V}\right) }$.
The U-duality group is $SO\left( 1,1\right) \times SO\left( 4,n_{V}\right)$
and the field strengths
transform under this group in the representations
 \bea
  F_2^\Lambda: &&       \left(
\mathbf{n}_{V}+\mathbf{4}\right) _{+\frac{1}{2}} \;,\nn\\
  H_3   : &&      \mathbf{1}_{\pm 1}\;.
 \eea
The coset representative $L_{\Lambda }^{~M }$, $\Lambda ,M
=1, \dots, 4+n_{V}$,  of $\frac{SO\left( 4,n_{V}\right) }{SO\left( 4\right)
\times SO\left( n_{V}\right) }$ sits in the $\left( \mathbf{4},\mathbf{n}%
_{V}\right) $ representation of the stabilizer $H=SO(4) \times
SO( n_{V}) \sim SU(2) _{L}\times SU(2)
_{R}\times SO(n_{V}) $, and satisfies the defining relations
\bea
L_{\Lambda}{}^M\,  \eta_{MN}\,L_{\Sigma}{}^N &=& \eta_{\Lambda \Sigma}
 \;,\qquad
L_{\Lambda}{}^M\,  \eta^{\Lambda \Sigma}  \,L_{\Sigma}{}^N ~=~
 \eta^{MN}
 \;,
\label{ll}
\eea
with the $SO(4,n_V)$ metric $\eta_{\Lambda \Sigma}$.
It is related to the vielbein ${\cal V}_\Lambda{}^M$ from (\ref{relations}) by
\bea
{\cal V}_\Lambda{}^M &=&
e^{-\phi/2}\,L_\Lambda{}^M
\;,
\eea
and its inverse is defined by
$L_M{}^\Lambda L_\Lambda{}^N=\delta_M^N$.
The Maurer-Cartan equations take the form
 \bea
P_{MN} &=&
L_M{}^\Lambda \,d_z L_{\Lambda N}
~=~L_M{}^\Lambda\, \partial_i L_{\Lambda N}\, dz^i
\;,
\label{MC11}
 \eea
where $P_{MN}$ is a symmetric
off-diagonal block matrix with non-vanishing entries
only in the $(4\times n_V)$-blocks.
Here and below we use $\delta_{MN}$ to raise and lower the indices $M,N$.

The solutions will be specified by the
electric and magnetic three-form charges $q$, $p$, and
the two-form charges $p^\Lambda$, $q_\Sigma$.
The quadratic and cubic U-duality invariants that can be built from these charges are
\bea
{\cal I}_2~=~ p q \;,\qquad
{\cal I}_3~=~\ft12 \eta_{\Lambda \Sigma}\,p^\Lambda p^\Sigma p\;,\qquad
{\cal I}_3'~=~\ft12 \eta^{\Lambda \Sigma}\,q_\Lambda q_\Sigma q\;. \label{uinv}
\eea
The central charges~(\ref{zcentral}), (\ref{centrald}) are given by
 \bea
 Z_{{\rm mag},M}&=& e^{-\phi/2} \,J_2\,  L_{\Lambda{}M}\,  p^\Lambda\;,
 \qquad\qquad
Z_{{\rm  el},M}=e^{\phi/2}\,J_2'\,  L_M{}^\Lambda\, q_\Lambda\;,
 \nn\\[.5ex]
 Z_{\pm}&=&  \frac1{\sqrt{2}}\,J_3\,  (e^{\phi}\, p\pm e^{-\phi}\, q)
 \;.
 \eea
  Using~(\ref{ll}),
  the U-duality invariants  (\ref{uinv}) can be rewritten in
  terms of the central charges as
\bea
\ft12\,(Z_+^2-Z_-^2)  &=& J_3^{2}\,{\cal I}_2\;,\nn\\
\ft1{2\sqrt{2}}\eta^{MN}\,Z_{\rm mag,M}  Z_{\rm mag,N} \,(Z_++Z_-) &=& (J_3 J_2^2)\, {\cal I}_3\;,\nn\\
\ft1{2\sqrt{2}}\eta^{MN}\,Z_{\rm el,M}  Z_{\rm el,N}\,(Z_+-Z_-)&=& (J_3 J'_2{}^{2})\, {\cal I}_3'   \;.
\label{uinvZ}
\eea
The effective potential
$V_{\rm eff}$ (\ref{Veff})
for this theory is given by
\bea
  V_{\rm eff}&=&
  \ft12  Z_{+}^2+ \ft12  Z_{-}^2 + \ft12 Z_{{\rm el},M}^{ 2}   +\ft12 Z_{{\rm mag},M}^{ 2}
  \;.
  \label{Veff_611}
\eea
From the Maurer-Cartan equations~(\ref{MC11}) one derives
 \bea
\nabla Z_{{\rm mag},M} &=&
-P_{MN}\, Z_{{\rm mag},N }-\ft12 P_\phi \,Z_{{\rm mag},M }
\;,\nn\\
\nabla Z_{{\rm el}, M}  &=&
P_{MN}\, Z_{{\rm el},N}+\ft12 P_\phi \,Z_{{\rm el},M}\;, \nn\\
  \nabla Z_{\pm}&=&P_\phi\, Z_{\mp}
 \;.
 \label{mc22}
 \eea
with $P_{\phi}= d\phi$. The attractor equations~(\ref{attractor})
thus translate into
\bea
   P_{MN}\,
  (Z_{{\rm el},M}  Z_{{\rm el},N} -
  Z_{{\rm mag},M} Z_{{\rm mag},N })
+
P_\phi\,   (   2Z_{+} Z_-   - \ft12 Z_{{\rm mag},M }^2+\ft12 Z_{{\rm el},M}^2 )
~\stackrel!{\equiv}~0\;.
\label{eq3}
  \eea
Splitting the index $M $ into $(A\dot A)=1, \dots, 4$, ($A,\dot A=1,2$)
(\textit{central charges sector}) and $I=5, \dots, (n_{V}+4)$ (\textit{matter
charges sector}), and using the fact that only the
components $P_{I,A\dot A}=P_{A\dot A,I}$ are non-vanishing, the
attractor equations can be written as
 \bea
 Z_{{\rm el},A\dot A}  Z_{{\rm el},I} -
  Z_{{\rm mag},A\dot A} Z_{{\rm mag},I }&=&0\;,\nn\\[1ex]
4Z_{+} Z_-   - Z_{{\rm mag},A\dot A }Z_{{\rm mag}}^{A\dot A }
+ Z_{{\rm el},A\dot A}Z_{{\rm el}}^{A\dot A} - Z_{{\rm mag},I }^2+ Z_{{\rm el},I}^2
&=&0\;.
\label{eq3A}
\eea
Indices $A,\dot A$ are raised and lowered by $\epsilon_{AB},\epsilon_{\dot A \dot B}$.
We will study the solutions of these equations,
their supersymmetry-preserving features,
and the corresponding moduli spaces.
BPS solutions correspond to the solutions of (\ref{eq3A})
satisfying
  \bea
 Z_{{\rm mag},I }~=~Z_{{\rm el},I}~=~ 0\;,
 \label{killeq}
 \eea
 as follows from the Killing spinor equation
 $\delta \lambda^I_A ~\sim~  T^I_ {\mu \nu} \gamma^{\mu\nu} \epsilon_{A}
 =0$ with $T^I_ {\mu \nu}$ the
 matter central charge densities.

Let us finally consider the  \textit{moduli space} of
the attractor solutions,
i.e.\  the scalar degrees of freedom which are \textit{not} stabilized by
the attractor mechanism at the classical level.
     For homogeneous scalar manifolds this space
is spanned by the vanishing eigenvalues of the Hessian matrix
$\nabla \nabla V_{\rm eff}$  at the critical point. Using the Maurer-Cartan equations (\ref{mc22}) one can write
$\nabla \nabla V_{\rm eff}$ at the critical point as
\bea
\nabla \nabla V_{\rm eff} &=& P_{I, A\dot{A}} P_{ J}{}^{A\dot{A}}\,(2\,  Z_{{\rm el},I}  Z_{{\rm el},J} +
  2\,  Z_{{\rm mag},I} Z_{{\rm mag},J })\nn\\
&&  + P^{I,A\dot{A} I} P^{I, B\dot{B}}\,(2\,  Z_{{\rm el}, A\dot{A}}  Z_{{\rm el},B\dot{B}} +
  2\,  Z_{{\rm mag}, A\dot{A}}  Z_{{\rm mag},B\dot{B}} )\nn\\
  &&+ P_\phi P_\phi\,  (   2Z_{+}^2+ 2Z_-^2   +\ft12 Z_{{\rm mag},M }^2+\ft12 Z_{{\rm el},M}^2 )\nn\\
  &&+2\,P_\phi P^{I, A\dot{A} }\,(  Z_{{\rm el},I}  Z_{{\rm el},A\dot{A}} +
    Z_{{\rm mag},I} Z_{{\rm mag},A\dot{A}})\nn\\[1ex]
    &=& H_{I A\dot{A},J B\dot{B} } P^{I,A\dot{A}} P^{J,B\dot{B}}+2\, H_{I A\dot{A},\phi} P^{I,A\dot{A}} P_{ \phi}+H_{\phi,\phi} P_{\phi} P_{ \phi}
    \;,\label{hessian}
\eea
which defines the Hessian symmetric matrix ${\bf H}$
with components $H_{I A\dot{A},JB\dot{B}}$, $H_{I A\dot{A},\phi}$,
$H_{\phi,\phi}$\,.  By explicit evaluation of the Hessian matrix for both BPS and non-BPS solutions
we will show that eigenvalues are always zero or positive implying the stability (at the classical level)
of the   solutions under consideration here.
We will now specify to the different near-horizon geometries and study the BPS and non-BPS
solutions of the attractor equations.


\subsection{$AdS_3\times S^3$}


Let us start with an $AdS_{3}\times S^{3}$ near-horizon geometry, in which only
the three-form charges (magnetic $p$ and electric $q$) are switched on
(\textit{dyonic black string}).
   There are no closed two-forms supported by this geometry
   and therefore two-form charges are not allowed.
   The near-horizon geometry ansatz can then be written as
    \bea
   ds^2 &=& r^2_{\!\rm AdS}\, ds^2_{\!\rm AdS_{3}}+ r^2_{\!S}\, ds^2_{S^{3}}\;,\qquad
    H_3 ~=~ p\, \alpha_{S^3}+ e\, \beta_{\!\rm AdS_3}
    \;.
    \eea
    The attractor equations~(\ref{eq3A}) are solved by
    \bea
    &&Z_{{\rm mag},M }=Z_{{\rm el}M}=Z_{-}=0  \;,\quad {\rm or~equivalently},    \label{solads3s31}\\
    &&Z_{{\rm mag},M }=Z_{{\rm el},M}=Z_+=0 \;.   \label{solads3s32}
    \eea
Solution (\ref{solads3s31}) has ${\cal I}_2>0$, whereas solution
(\ref{solads3s32}) has ${\cal I}_2<0$; they are both
$\frac{1}{4}$-BPS, and they are equivalent, because the considered
theory is non-chiral.

     Plugging the solutions  (\ref{solads3s31}) or   (\ref{solads3s32}) into (\ref{Veff_611})  one can write
   the effective potential at the horizon in the scalar
  independent form
  \bea
  V_{\rm eff} &=& \ft12 Z_{+}^2+\ft12 Z_{-}^2 = \ft12|  Z_{+}^2- Z_{-}^2 |   J_3^2 \,| {\cal I}_2|= \left({v_{{\rm\!AdS}_3}\over v_{\!S^3} }\right)\, |{\cal I}_2|
  \;,
    \eea
     in agreement with the claimed formula (\ref{sumads3s3}).  Extremizing   $F$ in $\vec{r}$, one finds the
     entropy  function and near-horizon AdS and sphere radii  (\ref{sumads3s3}).

      Now let us consider the moduli space of the solutions.
   Plugging (\ref{solads3s31}), (\ref{solads3s32}) into (\ref{hessian})
one finds that the only non-trivial component of the Hessian matrix
is \be H_{\phi\phi}=2 Z_+^2+2 Z_-^2=4 V_{\rm eff}>0 \;. \ee
Therefore, the Hessian matrix ${\bf H}$ for the $AdS_{3}\times
S^{3}$ solution has $4n_{V}$ vanishing eigenvalues and one strictly
positive eigenvalue, corresponding to the dilaton direction.
Consequently, the moduli space of non-degenerate attractors with
near-horizon geometry $AdS_{3}\times S^{3}$ is the quaternionic
symmetric manifold \be {\cal M}_{\rm BPS}~= \frac{SO\left(
4,n_{V}\right) }{SO\left( 4\right) \times SO\left( n_{V}\right) }
\;. \label{quaternionic} \ee This result is also evident from the
explicit form of the attractor solution
 $Z_-=0$: only the dilaton is
stabilized, while all other scalars are not fixed since the
remaining equations $Z_{{\rm el},M}=Z_{{\rm mag},M}=0$ are
automatically satisfied for $p^\Lambda=q_\Lambda=0$ .


\subsection{$AdS_3\times S^2\times S^1$}


For  solutions with near-horizon geometry
$AdS_3\times S^2\times S^1$,
there is no support for electric two-form charges and
therefore $e^\Lambda=0$. We set also the electric three-form
charge $e$   to zero otherwise no solutions are found.
The near-horizon ansatz becomes
  \bea
  ds^2 &=& r^2_{\!\rm AdS}\, ds^2_{\!\rm AdS_{3}}+ r^2_{\!S}\, ds^2_{S^{2}}+r_1^2 d\theta^2
  \;, \nn\\[.5ex]
 F^\Lambda_2 &=& p^\Lambda\, \alpha_{S^2}\;,\qquad
 H_3=  p\, \alpha_{S^2\times S^1}
 \;.
\eea The attractor equations  (\ref{eq3A}) admit two types of
solutions with non trivial central charges
    \bea
  {\rm BPS}:     &&
  Z_{+}=Z_{-} \;,
 \qquad
  Z_{{\rm mag},A\dot A }Z_{{\rm mag}}^{A\dot A }= 4 Z^2_{+} \;; \label{ads3s21}\\[1ex]
     \mbox{\rm non-BPS}:     &&
  Z_{+}=Z_{-}
 \qquad
  Z_{{\rm mag},I }^2= 4Z^2_{+} \;.
  \label{ads3s22}
  \eea
Plugging the solution into (\ref{uinvZ}) one finds the relation
\bea
  \left| J_2^2 J_3 \, {\cal I}_3 \right| &=& 2\sqrt{2}\,Z_+^3
    \;.
\eea
that allows us to write the effective potential~(\ref{Veff_611})
at the horizon in the scalar independent form
  \bea
  V_{\rm eff} &=& 3 Z_+^2~=~
  \ft32 \, \left| J_2^2 J_3 \, {\cal I}_3 \right|^{2\over 3} \label{ads3s2s111}
  \;,
  \eea
   with
   \be
  (J_2^2 J_3) ^{2\over 3}={v_D\over  ({\rm vol}_{S^2}^2  {\rm vol}_{S^2\times S^1}  )^{2\over 3}}
  ={v_{\!\rm AdS_3} \,v_{T}^{1\over 3}\,  \over v_{\!S}  }
\;,
 \ee
in agreement with our proposed formula (\ref{sumads3s2})  upon taking  ${\bf I}_3=  {\cal I}_3 $.
The black string central charge and the near-horizon radii follow from $\vec{r}$-extremization
of the entropy  function $F$ and are given by (\ref{sumads3s2}).
Note that the radius $r_1$ of the extra $S^1$ is not fixed by
the extremization equations. Besides this geometric modulus the solutions can be also deformed
by turning on Wilson lines for the vector field potentials $A^{\Lambda}_5=c^{\Lambda}$.
This is in contrast with the more familiar  case of black holes in $D=4,5$
where the near-horizon geometry is completely
fixed at the end of the attractor flow. As we shall see in the following, this will be always the case
for extremal black $p$-branes with $T^n$ factors where the ``geometric moduli" describing
the shapes ad volumes of the tori and  constant values of field potentials along
$T^n$ remain unfixed at the horizon.

Now, let us consider the moduli spaces of the two solutions.
The BPS solution  (\ref{ads3s21}) has
remaining symmetry $SO\left( 3\right) \times SO(n_{V})$,
because by using an $SO\left( 4\right) $ transformation
this solution can be recast in the form
\begin{equation}
Z_{{\rm mag},A\dot{A} }=2\,z \,\delta _{A 1} \delta _{\dot {A} 1} \;,
\quad\quad   Z_+=Z_-= z \;,
 \quad\quad Z_{{\rm el},M }=0
 \;.
  \label{2.1}
\end{equation}
Notice that both choices of sign satisfy the Killing spinor
relations (\ref{killeq}) and therefore correspond to
    supersymmetric solutions. Plugging (\ref{2.1}) into the Hessian matrix (\ref{hessian}) one finds
    \begin{equation}
{\bf H}= z^{2} \left(
\begin{array}{ccc}
8 \delta_{IJ} \delta_{A1} \delta_{B1} \delta_{\dot A1} \delta_{\dot B1}  &  & 0_{4n_{V}\times 1} \\[2ex]
0_{1\times 4n_{V}} &  & 6
\end{array}
\right) .  \label{tue-2}
\end{equation}
This matrix has $3n_{V}$ vanishing eigenvalues and $n_{V}+1$ strictly positive eigenvalues,
corresponding to the dilaton direction plus the $n_{V}$ directions~$P_{I,11}$ .
Consequently, the
{moduli space} of the BPS attractor solution~(\ref{ads3s21})
with near-horizon geometry $AdS_{3}\times S^{2}\times S^1$ is the symmetric
manifold
\be
{\cal M}_{\rm BPS}=\frac{SO( 3,n_{V}) }{SO(3) \times SO(n_{V}) }
\;.
\label{mbps22}
\ee
More precisely, the scalars along $P_{I,A\dot{A}}$
in the $\left( \mathbf{4},\mathbf{n}_{V}\right)$  of the group $H$
decompose with respect to the symmetry group $SO(3) \times SO(n_{V})$ as:
\begin{equation}
 \left( \mathbf{4},\mathbf{n}_{V}\right) \longrightarrow \underset{m^{2}=0}{%
\underbrace{\left( \mathbf{3},\mathbf{n}_{V}\right) }}\oplus \underset{%
m^{2}>0}{\underbrace{\left( \mathbf{1},\mathbf{n}_{V}\right) }},
\end{equation}
and only the $\left( \mathbf{1},\mathbf{n}_{V}\right) $
representation   is {massive}, together with
the dilaton. The $\left( \mathbf{3},\mathbf{n}_{V}\right) $ representation
 remains {massless}, and it contains all
the {massless Hessian modes} of the
attractor solutions.

The analysis of the moduli space for the  non-BPS solution follows closely that for the BPS one. Now
 the symmetry is $SO(4) \times SO(n_{V}-1)$ and using an $SO(n_V) $ transformation such a solution can
be recast as follows:
\begin{equation}
Z_{{\rm mag},I }=2 z\, \delta _{I 1}\;,  \quad\quad Z_+=Z_-= z\;,
\quad Z_{{\rm el},M}= Z_{{\rm mag},A\dot A}=Z_{{\rm el},A\dot
A}=0\;
 \quad
 \;.
  \label{2.1A}
\end{equation}
Plugging (\ref{2.1A}) into the Hessian matrix (\ref{hessian}), now one finds
    \begin{equation}
{\bf H}= z^{2} \left(
\begin{array}{ccc}
8 \delta_{A\dot A} \delta_{B\dot B}  \delta_{J1} \delta_{I1}   &  & 0_{4n_{V}\times 1} \\
&  &  \\
0_{1\times 4n_{V}} &  & 6
\end{array}
\right) .
\end{equation}
This Hessian matrix has $4(n_{V}-1)$ vanishing eigenvalues and $4+1$ strictly positive eigenvalues,
corresponding to the dilaton direction plus the $4$ $P_{1,A\dot A}$ directions.
Consequently, the \textit{moduli space} of the non-BPS attractor solution
  with near-horizon geometry $AdS_{3}\times S^{2}\times S^1$ is the symmetric
manifold
\be
{\cal M}_{\rm nonBPS}=\frac{SO(4,n_{V}-1) }{SO( 4) \times SO(
n_{V}-1) }
\;.
\label{mnonbps22}
\ee
More precisely, the scalars along
$P_{I,A\dot{A}}$  in the $\left( \mathbf{4},\mathbf{n}_{V}\right)$  of the group $H$
decompose with respect to the symmetry group $SO(4) \times SO(n_{V}-1)$ as:
\begin{equation}
 \left( \mathbf{4},\mathbf{n}_{V}\right) \longrightarrow \underset{m^{2}=0}{%
\underbrace{\left( \mathbf{4},\mathbf{n}_{V}\mathbf{-1}\right)
}}\oplus \underset{m^{2}>0}{\underbrace{\left(
\mathbf{4},\mathbf{1}\right) }},
\end{equation}
and only the $\left( \mathbf{4},\mathbf{1} \right) $
representation   is {massive},   together with
the dilaton. The $\left( \mathbf{4},\mathbf{n_V-1}\right) $ representation
  remains {massless}, and it contains all
the {massless Hessian modes} of the
attractor solution.

The BPS solution can be regarded as the intersection of one $\frac{1}{2}$%
-BPS black string (with $pq=0$) with one $\frac{1}{4}$-BPS black
$2$-brane (with $p^{\Lambda }p^{\Sigma }\eta _{\Lambda \Sigma }>0$).
The latter is described by the charge orbit $\frac{SO\left(
4,n_{V}\right) }{SO\left( 3,n_{V}\right) }$ \cite{AFMT1}. The moduli
space of the latter coincides with the moduli space of the whole
considered intersection, and it is given by Eq.~(\ref{mbps22}).

On the other hand, the non-BPS solution can be regarded as the
intersection of one $\frac{1}{2}$-BPS black string (with $pq=0$)
with one non-BPS black $2 $-brane (with $p^{\Lambda }p^{\Sigma }\eta
_{\Lambda \Sigma }<0$). The
latter is described by the charge orbit $\frac{SO\left( 4,n_{V}\right) }{%
SO\left( 4,n_{V}-1\right) }$ \cite{AFMT1}. The moduli space of the
latter coincides with the moduli space of the whole considered
intersection, and it is given by the quaternionic manifold of Eq.~(\ref{mnonbps22}).

A similar reasoning will be performed for the moduli spaces of the
attractor solutions of the maximal non-chiral $D=6$ supergravity in
Sect. \ref{Sect6}.


 \subsection{$AdS_2\times S^3\times S^1$}


For  solutions with $AdS_2\times S^3\times S^1$  near-horizon geometry,
there is no support for magnetic two-form charges and
therefore $Z_{{\rm mag}, M}=0$. The near-horizon ansatz becomes
  \bea
  ds^2 &=& r^2_{\!\rm AdS}\, ds^2_{\!\rm AdS_{2}}+ r^2_{\!S}\, ds^2_{S^{3}}+r_1^2 d\theta^2 \;,
  \nn\\[.5ex]
   F^\Lambda_2 &=& e^\Lambda\, \beta_{\!\rm AdS_2}\;,\qquad ~~~~  H_3=  e\, \beta_{\!\rm AdS_2\times S^1}
   \;.
   \eea
The fixed-scalar equations  (\ref{eq3}) admit two type of solutions
    \bea
  {\rm BPS}:     && Z_{{\rm mag},M}=  Z_{{\rm el},I}=0\;, \quad\quad ~    Z_{+}=-Z_{-}\;,
 \quad\quad
  Z_{{\rm el},A\dot A }Z_{{\rm el}}^{A\dot A }= 4 Z^2_{+} \;,\label{ads2s31}\\[1ex]
     \mbox{\rm non-BPS}:     && Z_{{\rm mag},M}= Z_{{\rm el},A\dot A}=0 \;,  \quad\quad ~    Z_{+}=-Z_{-}  \;,
 \quad\quad
  Z_{{\rm el},I }^2= 4Z^2_{+} \;.
  \label{ads2s32}
  \eea
 Now one finds
   \bea
     J_2^{'2} J_3'\,  {\cal I}_3'
= 2\sqrt{2}\, Z_+^3
    \;,
    \eea
  and  the effective potential (\ref{Veff_611})
    at the horizon can be written in the scalar
  independent form
        \bea
  V_{\rm eff} &=& 3 Z_+^2~=~ \ft32 \, | J_2^{'2} J_3'\,  {\cal I}_3' |^{2\over 3}
\;,
  \eea
   with
     \be
  (J_2^{'2} J_3') ^{2\over 3}={   ({\rm vol}_{\!\rm AdS_2}^2  {\rm vol}_{\!\rm AdS_2\times S^1}  )^{2\over 3}\over v_D}
  ={v_{\!\rm AdS_2} \over  v_{T}^{1/3}\, v_{\!S}  }
\;,
 \ee
  in agreement with the proposed formula (\ref{sumads2s3}) upon taking ${\bf I}_3=  {\cal I}_3' $.
  Extremizing $F$ in the radii $\vec{r}$ one finds the result~(\ref{sumads2s3})
  for the black hole entropy and AdS and sphere radii.
  Again, the radius $r_1$ of the extra $S^1$ is not fixed by
the extremization equations.
      The analysis of the moduli spaces   follows {\it mutatis mutandis}
      that of the $AdS_3\times S^2$ attractors
      (replacing magnetic by electric charges)
     and the results are again given by the symmetric manifolds
     (\ref{mbps22}) and (\ref{mnonbps22}).


\section{${\cal N}=(2,2)$ in $D=6$}\label{Sect6}


\subsection{${\cal N}=(2,2)$, $D=6$ Supersymmetry
Algebra}

The maximal $\left( 2,2\right) $, $D=6$ Poincar\'{e} supersymmetry
algebra has Weyl pseudo-Majorana supercharges and
$\mathcal{R}$-symmetry $USp\left( 4\right) _{L}\times USp\left(
4\right) _{R}$ ($USp\left( 4\right) = Spin(5) $). Its central
extension reads as follows (see \textit{e.g.}\
\cite{Strathdee:1986jr,VanProeyen:1999ni,Townsend:1995gp})
\begin{eqnarray}
\left\{ \mathcal{Q}_{\gamma }^{A},\mathcal{Q}_{\delta
}^{B}\right\} &=&\gamma _{\gamma \delta }^{\mu }Z_{\mu }^{\left[
AB\right] }+\gamma
_{\gamma \delta }^{\mu \nu \rho }Z_{\mu \nu \rho }^{\left( AB\right) }; \\
\left\{ \mathcal{Q}_{\dot{\gamma}}^{\dot A},\mathcal{Q}_{\dot{\delta}%
}^{\dot B}\right\}  &=&\gamma _{\dot{\gamma}\dot{\delta}\ }^{\mu
}Z_{\mu }^{\left[ \dot A\dot B \right] }+\gamma _{\dot{\gamma}\dot{%
\delta}\ }^{\mu \nu \rho }Z_{\mu \nu \rho }^{\left( \dot A\dot B \right) }; \\
\left\{ \mathcal{Q}_{\gamma
}^{A},\mathcal{Q}_{\dot{\delta}}^{\dot A}\right\}  &=&C_{\gamma \dot{\delta}}Z^{A\dot A}+\gamma _{\gamma \dot{%
\delta}\ }^{\mu \nu }Z_{\mu \nu }^{A\dot A},
\end{eqnarray}
where $A,\dot A=1, \dots, 4$, so that the (L,R)-chiral supercharges are $%
SO\left( 5\right) _{\left( L,R\right) }$-spinors.

Strings can be dyonic, and they are in the antisymmetric traceless
$\left( \mathbf{5},\mathbf{1}\right) +\left(
\mathbf{1},\mathbf{5}^{\prime }\right) $ of the
$\mathcal{R}$-symmetry group. They are embedded in the
$\mathbf{10}$ of the $U$-duality group $SO\left( 5,5\right) $. On
the other hand, black
holes and their magnetic duals (black $2$-branes) sit in the $\left( \mathbf{%
4},\mathbf{4}^{\prime }\right) $ of $USp\left( 4\right) _{L}\times
USp\left(
4\right) _{R}$, and they are embedded in the chiral spinor repr. $\mathbf{16}%
_{\left( L\right) }$ of $SO\left( 5,5\right) $.

In our analysis, the corresponding central charges are denoted
respectively
by $Z_{a}$ and $Z_{\dot a}$ ($a,\dot a=1, \dots, 5$) for dyonic strings, and by $%
Z_{{\rm el},A \dot{A}}$ and $Z_{{\rm mag},A \dot{A}}$ for black holes and
their magnetic duals.

\subsection{${\cal N}=(2,2)$, $D=6$ Supergravity}

The maximal ${\cal N}=(2,2)$ supergravity in $D=6$
dimensions~\cite{Tanii:1984zk} has bosonic field content given by
the graviton, 25 scalar fields, 16 vectors and 5 two-form fields.
Under the global symmetry group $SO(5,5)$ these fields organize as
 \bea
 {\cal V}_{I}{}^M ~=~\left(
 \begin{array}{cc}
 {\cal V}_{m}{}^a&{\cal V}_{m}{}^{\dot{a}}\\
 {\cal V}^{m}{}^a&{\cal V}^{m}{}^{\dot{a}}
 \end{array}
 \right)
 :&&
{SO(5,5)\over SO(5)\times SO(5)}\quad\quad
\begin{array}{l}
I,M=1,\dots10\,,\\
a,\dot a,m=1,\dots, 5\,,
\end{array}
\nn\\[1ex]
  F_2^{\Lambda}: &&     {\bf 16}    \quad\quad \Lambda=1, \dots,16\,, \nn\\[.5ex]
 \{ H_{3+}^a,~H_{3-}^{\dot a}\} : && {\bf 10}
  \quad\quad  a,\dot a=1, \dots, 5
  \;.
 \eea
 In particular, the scalar coset space is parametrized by the vielbein ${\cal V}_{I}{}^M$
 evaluated in the vector representation ${\bf 10}$ of $SO(5,5)$,
 satisfying the defining relations
 \be
{\cal V}_{I}{}^{a}{\cal V}_{J}{}^{a}-{\cal V}_{I}{}^{\dot{a}} {\cal V}_{J}{}^{\dot{a}}
=\eta_{IJ}\equiv\left(
 \begin{array}{cc}
 0&1\\
 1&0
 \end{array}
 \right)
 \;,\quad
 {\cal V}_{m}{}^{M}{\cal V}^m{}^{N}+{\cal V}^m{}^{M} {\cal V}_{m}{}^{N}
=\eta^{MN}\equiv\left(
 \begin{array}{cr}
 1&0\\
 0&-1
 \end{array}
 \right)
\;,
\nonumber
\ee
i.e.\ the splits of basis ${\cal V}_{I}{}^M\rightarrow ({\cal V}_{m}{}^M,{\cal V}^m{}^M)$
and ${\cal V}_{I}{}^M\rightarrow ({\cal V}_{I}{}^a,{\cal V}_I{}^{\dot{a}})$
refer to the decompositions
$SO(5,5)\rightarrow GL(5)$ and $SO(5,5)\rightarrow SO(5)\times SO(5)$,
respectively.
They are relevant for splitting the two-forms into electric and magnetic
potentials and for coupling them to the fermionic fields, respectively.
The scalar coset space can equivalently be described by a  scalar
vielbein ${\cal V}_\Lambda{}^{A\dot{A}}$ ($A,\dot A=1, \dots, 4$)
evaluated in the ${\bf 16}$  spinor representation of $SO(5,5)$.
The Maurer-Cartan equations are given by \bea \nabla {\cal
V}_{I}{}^a =  -P_{a\dot{a}} {\cal V}_{I}{}^{\dot{a}} \;,\qquad
\nabla {\cal V}_{I}{}^{\dot{a}} =-P_{a\dot{a}} {\cal V}_{I}{}^{a}
\;,\qquad \nabla {\cal V}_\Lambda{}^{A\dot{A}} = -\ft12
P_{a\dot{a}} \, \gamma_a^{AB}\gamma_{\dot a}^{\dot{A}\dot{B}}\,
{\cal V}_\Lambda{}_{B\dot{B}} \;, \label{MC_622} \eea with the
$SO(5)\times SO(5)$ Gamma matrices $\gamma_a^{AB}$, $\gamma_{\dot
a}^{\dot{A}\dot{B}}$, and the vector and spinorial indices raised
and lowered by the $SO(5)$ invariant symmetric tensors
$\delta_{ab}$, $\delta_{\dot a\dot b}$ and antisymmetric tensors
$\Omega_{AB}$, $\Omega_{\dot A\dot B}$, respectively.

The Lagrangian involves the 5 two-forms $B^m$, whose field
strengths are related to the selfdual $H_{3+}^a$ and antiselfdual
$H_{3-}^{\dot a}$ by \bea dB^m &=& H^m ~\equiv~ {\cal V}^m{}^a
H_{3+}^a +  {\cal V}^m{}^{\dot{a}} H_{3-}^{\dot{a}}\;. \eea
Electric and magnetic three-form charges combine into an $SO(5,5)$
vector $Q_I = (p^m, q_m)$. The quadratic and cubic U-duality
invariants of charges are given by
\bea {\cal I}_2 =\ft12
\eta^{IJ} Q_I Q_J
\;,\quad {\cal I}_3  =\ft{1}{2\sqrt{2}}
(\Gamma^I)_{\Lambda \Sigma}\,Q_I\,p^\Lambda p^\Sigma \;,\quad
{\cal I}_3'  =\ft{1}{2\sqrt{2}} (\Gamma^I)^{\Lambda
\Sigma}\,Q_I\,q_\Lambda q_\Sigma  \label{uinv22} \eea with the
$SO(5,5)$ Gamma matrices $(\Gamma^I)^{\Lambda \Sigma}$,
$(\Gamma^I)_{\Lambda \Sigma}$.

The central charges~(\ref{zcentral}), (\ref{centrald}) are defined
as \bea Z_{{\rm mag}}^{A\dot{A}} &=& J_2\, {\cal
V}_{\Lambda}{}^{A\dot{A}}   \,p^\Lambda\;, \qquad Z_{{\rm
el}}^{A\dot{A}} ~=~ J_2'\, ({\cal V}^{-1})^{A\dot{A} \,\Lambda   }
\,q_\Lambda\;,
\nonumber\\
Z_a &=& J_3\,({\cal V}^{-1})_a{}^I\,Q_I \;,\qquad
Z_{\dot{a}} ~=~  J_3\,({\cal V}^{-1})_{\dot{a}}{}^I\,Q_I
\;.
\eea
In terms of these central charges one can
 rewrite the U-duality invariants (\ref{uinv22}) as
\bea
2\, J_3^2\, {\cal I}_2 &=& Z_{a}^2-Z_{\dot a}^2 \;,\nn\\
 2\sqrt{2}\,(J_2^2 J_3)\, {\cal I}_3 &=&  Z_{\rm mag}^{A\dot{A}}\,Z_{\rm mag}^{B\dot{B}} \,(Z_a \gamma^a_{AB}\Omega_{\dot{A}\dot{B}}+Z_{\dot a} \Omega_{AB}\gamma^{\dot a}_{\dot{A}\dot{B}})\;,\nn\\
 2\sqrt{2}\,(J'_2{}^{2} J_3)\, {\cal I}'_3\ &=&  Z_{{\rm
el},A\dot{A}}\,Z_{{\rm el},B\dot{B}} \,(Z_a
\gamma^a_{AB}\Omega_{\dot{A}\dot{B}} -Z_{\dot a}
\Omega_{AB}\gamma^{\dot a}_{\dot{A}\dot{B}})\;. \label{uinvZ22}
\eea The {intersecting-branes effective potential} $V_{\rm eff}$
for the considered theory is defined as \be V_{\rm eff}= \ft12
Z_{a}^2+\ft12 Z_{\dot{a}}^2 +
 \ft12 Z_{{\rm el},A\dot{A}}Z_{{\rm el}}^{A\dot{A}}
+  \ft12 Z_{{\rm mag},A\dot{A}}Z_{{\rm mag}}^{A\dot{A}} \;.
\label{veff22} \ee The Maurer-Cartan equations (\ref{MC_622})
imply \bea \nabla {Z}_a &=& P_{a\dot{a}} Z_{\dot{a}} \;,\qquad
\nabla Z_{\dot{a}} =  P_{a\dot{a}} Z_{a}\nn\\
\nabla Z_{\rm el}^{A\dot{A}} &=&
 \ft12  P_{a\dot{a}} \, \gamma_a^{AB}\gamma_{\dot a}^{\dot{A}\dot{B}}\,
Z_{{\rm el},B\dot{B}} \;,\qquad \nabla Z_{\rm mag}^{A\dot{A}} =-
  \ft12 P_{a\dot{a}} \, \gamma_a^{AB}\gamma_{\dot a}^{\dot{A}\dot{B}}\,
Z_{{\rm mag},B\dot{B}} \;. \eea Thus the extremization equations
take the form
 \bea
\nabla V_{\rm eff}&=&
 Z_{a}\,\nabla Z_{a}+ Z_{\dot{a}}\,\nabla Z_{\dot{a}}+
 Z_{{\rm el},A\dot{A}}   \,\nabla Z_{\rm el}^{A\dot{A}}
+  Z_{{\rm mag},A\dot{A}} \,\nabla Z_{\rm mag}^{A\dot{A}}
\nonumber\\
&=& \Big( 2Z_{a} Z_{\dot{a}} + \ft12 \gamma^a_{AB}\gamma^{\dot
a}_{\dot{A}\dot{B}} Z_{\rm el}^{A\dot{A}}\,Z_{\rm el}^{B\dot{B}} -
\ft12 \gamma^a_{AB}\gamma^{\dot a}_{\dot{A}\dot{B}} Z_{\rm
mag}^{A\dot{A}}\,Z_{\rm mag}^{B\dot{B}}\Big)\, P_{a\dot{a}}
~\stackrel{!}\equiv~ 0 \;. \label{eq6} \eea
 The Hessian matrix at the horizon can written as
  \bea
\nabla \nabla V_{\rm eff}&=&
2P_{a\dot{a}} P_{b\dot{b}} \, \Big\{ \delta_{ab}Z_{\dot{a}}Z_{\dot{b}}+\delta_{\dot{a}\dot b}%
Z_{a}Z_{b}\nn\\
&&  \qquad ~~~~~~~~~+\ft14 (\gamma^{a}\gamma^{b})_{A}{}^B
(\gamma^{\dot a}\gamma^{\dot b})_{\dot{A}}{}^{\dot{B}}
\Big( Z_{\rm el}^{A \dot{A}}Z_{{\rm el},B \dot{B}%
}+Z_{\rm mag}^{A \dot{A}}Z_{{\rm mag},B \dot{B}}\Big) \Big\} \;.
\label{hes2} \eea


\subsection{$AdS_3\times S^3$}


   The analysis of $D=6$ maximal supersymmetric supergravity solutions follows the same steps as in the half-maximal
   case with minor modifications.
   We start from the ansatz
    \bea
   ds^2 &=& r^2_{\!\rm AdS}\, ds^2_{\!\rm AdS_{3}}+ r^2_{\!S}\, ds^2_{S^{3}} \;,\nn\\
    H^m_3 &=& p^m\, \alpha_{S^3}+ e^m\, \beta_{\!\rm AdS_3}\;,
    \eea
for the $AdS_3\times S^3$ near-horizon geometry.
    The fixed scalar equation  (\ref{eq6}) admits the two solutions
    \bea
      &&
   Z_{{\rm mag},A \dot A }=Z_{{\rm el},A\dot A }=Z_{a}=0\;, \nn\\
    &&
   Z_{{\rm mag},A\dot A }=Z_{{\rm el},A\dot A }=Z_{\dot a}=0\;,
    \eea
which are both supersymmetric.
  Combining this with (\ref{uinvZ22}) one can write the effective potential at the horizon in the scalar
  independent form
  \bea
  V_{\rm eff} &=&  \ft12( Z_{a}^2+ Z_{\dot a}^2)=\ft12 |Z_{a}^2- Z_{\dot a}^2|= J_3^2 \, |{\cal I}_2|
\;.
  \eea
Again the effective potential is given by the general formula (\ref{sumads3s3})
but now ${\bf I_2}={\cal I}_2$ is given by the quadratic
invariant~(\ref{uinv22})  of $SO(5,5)$.
Similarly, $\vec{r}$-extremization of the entropy function shows that
the sphere and AdS radii and the black string central charges are given by  (\ref{sumads3s3})
in terms of the $SO(5,5)$ invariant ${\cal I}_2$.

   Let us consider the moduli space of these solutions. The two solutions are equivalent and
   we can focus on the $Z_a=0$ case.  Using an $SO(5)$ rotation this solution can be recast in
 the form $Z_{\dot a}=z \delta_{\dot a, 1}$. The symmetry group leaving this solution
invariant is $SO(5,4)$. The moduli space is hence given by the
quotient of this group  by its maximal compact subgroup $SO(5)\times
SO(4)$, i.e. (\cite{Ferrara:1997ci,Lu:1997bg,ADFL6})
 \bea {\cal M}_{\rm BPS}&=&\frac{SO(5,4)
}{SO(5) \times SO(4) } \;. \eea Alternatively, the same conclusion
can be reached by evaluating  the Hessian~(\ref{hes2}) at the
solution
 \bea
\nabla\nabla V_{\rm eff}&=& 2\, z^2 P_{a\dot{1}}P_{a\dot{1}}
\;,
\eea
 one finds $5$ strictly positive eigenvalues. More precisely, the $\left(
\mathbf{5},\mathbf{5}\right)$ scalars decompose in terms of $SO(5) \times SO(4)$
as
 \begin{equation}
 \left( \mathbf{5},\mathbf{5}\right) \longrightarrow \underset{m^{2}=0}{%
\underbrace{\left( \mathbf{5},\mathbf{4}\right) }}\oplus \underset{m^{2}>0}{%
\underbrace{\left( \mathbf{5},\mathbf{1}\right) }},
\label{mon-1-decomp}
\end{equation}
 with the $({\bf 5},{\bf 4})$ components along $P_{a,\dot{b}>\dot 1}$
 spanning the moduli space of the solution.

The story goes the same way   for the solution with $Z_{\dot a} = 0$
which has moduli space ${\cal M}_{\rm BPS}=\frac{SO(4,5) }{SO(4)
\times SO(5) }$.  The two solutions are equivalent and they both
preserve the same amount of supersymmetry (namely the minimal one:\
$\ft18$-BPS).
Actually, they can be interpreted as the \textit{%
supersymmetry}
 \textit{uplift} of the two distinct $\frac{1}{4}$-BPS solutions (given by Eqs. (\ref{solads3s31})
 and (\ref{solads3s32}))
   of  the half-maximal  $D=6$ supergravity
coupled to $n_{V}=4$ vector multiplets.


 \subsection{$AdS_3\times S^2\times S^1$}


 The ansatz for this near-horizon geometry is
  \bea
  ds^2 &=& r^2_{\!\rm AdS}\, ds^2_{\!\rm AdS_{3}}+ r^2_{\!S}\, ds^2_{S^{2}}+r_1^2 d\theta^2\;, \nn\\
   F^A_2 &=& p^A\, \alpha_{S^2}\quad\quad ~~~~  H_3^m=  p^m\, \alpha_{S^2\times S^1}\;.
   \eea

   Notice that a magnetic string corresponds to the $SO\left( 5,5\right) $
invariant constructed with the $10$-dimensional vector $\left(
p^{m},0\right) $ having vanishing norm. This is the
$\frac{1}{2}$-BPS
constraint for a $D=6$ string configuration, derived in \cite{Ferrara:1997ci}%
.

    The solutions of the fixed-scalar equations  (\ref{eq6}) on this background
    can be written up to an $SO(5)\times SO(5)$ rotation as\footnote{
    The explicit form of the solution clearly depends on the particular form of
    $SO(5)$ gamma-matrices considered. In our conventions, this
    choice of $Z_{\rm mag}^{A \dot{A }}$
    induces a matrix $\gamma^a_{AB}\gamma^{\dot a}_{\dot{A}\dot{B}}
Z_{\rm mag}^{A\dot{A}}\,Z_{\rm mag}^{B\dot{B}}$ which has only one
non-vanishing entry.}
    \bea
      Z_a &=& z \delta_{a1} \;, \qquad  Z_{\dot a}~=~z \delta_{\dot a1} \;,
     \nn\\
      Z_{{\rm el}, A \dot A }&=&0 \;, \qquad
 ~~~~ Z_{\rm mag}^{A \dot{A }}~=~ \sqrt{2}\,{\rm diag}( z,  z,0,0) \;.
 \label{zsol22}
    \eea
Using (\ref{uinvZ22}) one can express $z$ in terms of the cubic U-invariant~(\ref{uinv22})
     \be
     (J_2^2 J_3)\, {\cal I}_3 ~=~ 2\sqrt{2}  \, z^3 \;.
     \label{iz}
     \ee
        Combining (\ref{veff22}), (\ref{zsol22}), (\ref{iz}),
        one finally writes  the effective potential  in the scalar independent form
                  \bea
  V_{\rm eff} &=&
3\,  z^2
~=~ \ft32 \, \left| J_2^2 J_3 \,  {\cal I}_3 \right|^{2/3}
~=~\ft32\, \frac{v_{\!\rm AdS_3} \,v_{T}^{1/3}}{v_{\!S}  }  \, |{\cal I}_3 |^{2/3}
\;.
  \eea
 Like in the half-maximal case, the effective potential, the black string central charge
 and the near-horizon geometry
  are given by the general formulas   (\ref{sumads3s2}) but now in terms  of
   the  $SO(5,5)$ cubic invariant  ${\bf I}_3={\cal I}_3$\,.
   Again, the radius $r_1$ of the extra $S^1$ is not fixed by
the extremization equations.

  Let us consider the moduli space of this attractor.
  The symmetry of the solution  is $SO(4,3)$, which is
  the subgroup of $SO(5,5)$   leaving invariant
 (\ref{zsol22}). To see this we notice that
 $SO(4,3)$ is the maximal subgroup of $SO(5,5)$
  under which the decompositions
 of both the vector  and the spinor representations  of $SO(5,5)$ contain
 a singlet
 \bea
 {\bf 10} &=&{\bf 7} \oplus  3\cdot {\bf 1} \;,  \nn\\
  {\bf 16}&=&{\bf 8} \oplus {\bf 7}  \oplus {\bf 1}\;.
 \label{so7dec}
 \eea
 The moduli space is then given by the quotient of the symmetry group by its maximal compact subgroup
 \bea
{\cal M}_{\rm BPS} &=&\frac{SO( 4,3) }{SO(4) \times SO(3) }
\;.
\label{so43}
\eea
 More precisely, decomposing the scalar components $P_{a\dot a}$ under $SO(4)\times SO(3)$ one
 finds
  \be
  \left( \mathbf{5},\mathbf{5}\right) \longrightarrow \underset{m^{2}=0}{%
\underbrace{\left( \mathbf{4},\mathbf{3}\right) }}\oplus \underset{m^{2}>0}{%
\underbrace{\left( 2\cdot \left( \mathbf{4},\mathbf{1}\right) \oplus
\left(
\mathbf{1},\mathbf{3}\right) \oplus 2\cdot \left( \mathbf{1},\mathbf{1}%
\right) \right) }}. \ee
 This can be confirmed by explicitly evaluating
the Hessian~(\ref{hes2}) at this extremum. As a result one finds 12
vanishing and 13 strictly positive eigenvalues.

The moduli space in (\ref{so43}) can be understood in terms of orbits of $%
\frac{1}{2}$-BPS strings and $\frac{1}{4}$-BPS black holes \cite
{Ferrara:1997ci,Lu:1997bg,ADFL6}. Indeed, the $U$-invariant
$\mathcal{I}_{3}$ can be considered as an intersection of a
$\frac{1}{2}$-BPS string with supporting charge orbit
$\frac{SO\left( 5,5\right) }{SO\left( 4,4\right) \times
_{s}\mathbb{R}^{8}}$ and of a $\frac{1}{4}$-BPS black hole with
supporting charge orbit $\frac{SO\left( 5,5\right) }{SO\left(
4,3\right) \times _{s}\mathbb{R}^{8}}$ \cite{Lu:1997bg}. The common
stabilizer of the
charge vectors $\mathbf{10}$ and $\mathbf{16}$ of the $D=6$ $U$-duality $%
SO(5,5)$ is $SO\left( 4,3\right) $. Indeed, we find that the
resulting
moduli space of the considered intersecting configuration is given by Eq.~(%
\ref{so43}). This is also what expected by the supersymmetry uplift
of the BPS moduli space of the half-maximal $\left( 1,1\right) $
theory to maximal $\left( 2,2\right) $ supergravity.


\subsection{$AdS_2\times S^3\times S^1$}


 The near-horizon geometry ansatz is
  \bea
  ds^2 &=& r^2_{\!\rm AdS}\, ds^2_{\!\rm AdS_{2}}+ r^2_{\!S}\, ds^2_{S^{3}}+r_1^2 d\theta^2 \;,\nn\\
   F^A_2 &=& e^A\, \beta_{\!\rm AdS_2}\;,\quad\quad ~~~~  H_3^m=  e^m\, \beta_{\!\rm AdS_2\times S^1}
   \;.
   \eea
The computation of the effective potential proceeds as for the $AdS_3\times S^2\times S^1$ case
replacing magnetic by electric charges. The final result read
          \bea
  V_{\rm eff} &=&
 \ft32 \, \left| J_2^{'2} J_3 \, {\cal I}_3' \right|^{2\over 3}
 ~=~\ft32 \,{v_{\!\rm AdS_2}\over v_{S^3}  \,v_{T}^{1/3}}  \,| {\cal I}'_3|
 \;.
  \eea
    Extremizing the entropy  function $F$ in the radii $\vec{r}$ one confirms that
     the AdS, sphere radii and the black hole entropy are given
     again by the general formulae (\ref{sumads2s3}) with
     ${\bf I_3}={\cal I}_3'$  the magnetic $SO(5,5)$ cubic invariant.
       The analysis of the moduli space is identical to that of
       $AdS_3\times S^2\times S^1$ case and
       the result is again given by (\ref{so43}).


\section{Maximal $D=7$}\label{Sect7}


\subsection{$\mathcal{N}=2$, $D=7$ Supersymmetry
Algebra}

The maximal $\mathcal{N}=2$, $D=7$ Poincar\'{e} supersymmetry
algebra has
pseudo-Majorana supercharges and $\mathcal{R}$-symmetry $USp\left( 4\right) $%
. Its central extension reads as follows (see \textit{e.g.}\
\cite{Strathdee:1986jr,VanProeyen:1999ni,Townsend:1995gp})
\begin{equation}
\left\{ \mathcal{Q}_{\gamma }^{A},\mathcal{Q}_{\delta
}^{B}\right\} =C_{\gamma \delta }Z^{\left( AB\right) }+\gamma
_{\gamma \delta }^{\mu }Z_{\mu }^{\left[ AB\right] }+\gamma
_{\gamma \delta }^{\mu \nu }Z_{\mu \nu }^{\left[ AB\right]
}+\gamma _{\gamma \delta }^{\mu \nu \rho }Z_{\mu \nu \rho
}^{\left( AB\right) }
\end{equation}
where $A=1, \dots, 4$, so that the supercharges are $SO\left( 5\right)
$-spinors. The "trace" part of $Z_{\mu }^{\left[ AB\right] }$ is
the momentum $P_{\mu }\Omega ^{AB}$, where $\Omega ^{AB}$ is the
$4\times 4$ symplectic metric.

Black holes and their magnetic dual (black $3$-brane) central extensions
 $Z^{\left( AB\right) }, Z_{\mu \nu \rho
}^{\left( AB\right) }$  sit in the $\mathbf{%
10}$ of the $\mathcal{R}$-symmetry group, and they are embedded in the $%
\mathbf{10}$ (and $\mathbf{10}^{\prime }$) of the $U$-duality group $%
SL\left( 5,\mathbb{R}\right) $. Thus, they correspond to the decomposition $%
\mathbf{10}^{\left( \prime \right) }\longrightarrow \mathbf{10}$ of $%
SL\left( 5,\mathbb{R}\right) $ into $SO\left( 5\right) $.

On the other hand, black strings and their magnetic dual (black
$2$-brane) central extensions  $Z_{\mu }^{\left[ AB\right]}, Z_{\mu \nu }^{\left[ AB\right]
}$  sit in the $\mathbf{5}$ of $USp\left( 4\right) $, and
they are embedded in the $\mathbf{5}^{\prime }$ (and $\mathbf{5}$)
of the $U$-duality group. Thus, they correspond to the decomposition
$\mathbf{5}^{\left( \prime \right) }\longrightarrow \mathbf{5}$ of
$SL\left( 5,\mathbb{R}\right) $ into $SO\left( 5\right) $.

In our analysis, the corresponding central charges are denoted by $%
Z_{\rm el}^{mn}$ and $Z_{\rm mag}^{mn}$ ($m,n=1, \dots, 5$ are $SO\left(
5\right) $
indices) for black holes and their magnetic duals, and by $Z_{\rm el}^{m}$ and $%
Z_{\rm mag}^{m}$ for black strings and their magnetic duals.

\subsection{$\mathcal{N}=2$, $D=7$ Supergravity}

The global symmetry group of maximally supersymmetric $D=7$
supergravity~\cite{Sezgin:1982gi} is $SL(5,\mathbb{R})$. The bosonic
field content comprises the graviton, 14 scalars, 10 vectors and 5
two-form fields. Under the U-duality group $SL(5,\mathbb{R})$ these
organize as \bea
 {\cal V}^i{}_m ;~&&
{SL(5,\mathbb{R})\over SO(5)}\quad\quad i,m=1,\dots 5\,,  \nn\\
  F_2^{[ij]}: &&  {\bf 10'} \;, \nn\\
 H_{3 i}: && {\bf 5}
 \;.
\eea
The corresponding charges will be denoted by $p^{ij}, q_{ij}, p_i, q^i$.
 For near-horizon geometries  $AdS_2\times S^3\times T^2$
 and $AdS_3\times S^2\times T^2$,
 there are two independent electric and magnetic three-cycles, respectively,
 depending on which of the two circles of $T^2=S^1_1\times S^1_2$ is
 part of the cycle. The
 corresponding three-charges  will be denoted by $p_{ia}, q^i_r$ with $a,r=1,2$.


\subsection{$AdS_3\times S^3\times S^1$}


We start from the ansatz
\bea
ds^2 &=& r_{\!\rm AdS}^2\, ds^2_{\!\rm AdS_3} + r_{\!S}^2 \, ds^2_{S^3} +
r_1^2 d\theta_1{}^2
\;,
\nonumber\\[1ex]
H_{3 i}&=&
e_i\,\beta_{\!\rm AdS_3}
+p_i\,\alpha_{S^3}
\;,
\eea
for the near-horizon geometry.
The central charges and the relevant quadratic
U-duality invariant  are given by
 \bea
Z_{{\rm mag},m} &=& J_3\, {\cal V}^j{}_m\,p_{j}
\;,\qquad
Z_{{\rm el},\,m} ~=~ J'_3\, ({\cal V}^{-1})^m{}_i\,q^i\;,\nn\\[.5ex]
   {\cal I}_2 &=& q^{i}p_{i} ~=~ Z_{{\rm mag},i} Z_{\rm el}^i \, (J_3 J_3' )^{-1}
   \;,
 \eea
with $J_3=v_T J_3'=\left(  {v_{\rm\!AdS_3} v_T \over    v_{\!S^3} }\right)^{1\over 2} $.
  The effective potential can be written as
\bea
V_{\rm eff}&=&
\ft12Z_{{\rm mag}}^{m} Z_{{\rm mag}}^{m} +
\ft12Z_{{\rm el},\,m} Z_{{\rm el},\,m}
\;.
\eea
Using the Maurer-Cartan equations, we obtain
\bea
\nabla Z_{{\rm mag}}^{m}  =  Z_{{\rm mag}}^{n}  P_{mn}
\;,\qquad
\nabla Z_{{\rm el},\,m}  = -Z_{{\rm el},\,n}  P_{mn}
\;,
\eea
with $P_{mn}$ a symmetric and traceless matrix ($P_{mm}=0$).
Here, indices $m, n$ are raised and lowered with $\delta_{mn}$.
For the variation of the effective potential we thus obtain
\bea
\nabla V_{\rm eff} &=&
\Big(
Z_{{\rm mag}}^{m} Z_{{\rm mag}}^{n} -
Z_{{\rm el},\,m} Z_{{\rm el},\,n}
\Big)\, P_{mn}
~\stackrel{!}\equiv~ 0
\;,
\label{eq7_3}
\eea
Equation~(\ref{eq7_3})  is solved by
\be
Z_{{\rm mag}}^{m}=\pm Z_{{\rm el},\,m}
\;.
\ee
In this case we find
\bea
V_{\rm eff}\Big|_{\nabla V_{\rm eff}\equiv0} &=&
 J_3{}J'_3\,|{\cal I}_2|~=~\frac{v_{\!\rm AdS_3}}
{ v_{\!S^3} }
\;|{\cal I}_2|
\;,
 \eea
 in agreement with (\ref{sumads3s3}).
Extremization of $F$ w.r.t.\
the radii yields the black string central charge and near-horizon geometry
(\ref{sumads3s3} ) with
 ${\bf I_2}={\cal I}_2$ the $SL(5,\mathbb{R})$ quadratic invariant.
Notice  that  this solution can be thought of as the $D=7$ lift of
the $AdS_3\times S^3$ solution studied in the last section. The
radius $r_1$ of the additional $S^1$ is not fixed by the attractor
equations.

Finally  let us consider the moduli space of this black string
solution. For this purpose we notice that upon $SO(5)$ rotation the
solution can written in the form \be Z_{{\rm mag}}^{m}~=~\pm Z_{{\rm
el},\,m}~=~z \delta_{m1} \;. \ee This form is clearly invariant
under  $SL(4,\mathbb{R})$ rotations. The moduli space can then be
written as \bea {\cal M}_{\rm BPS}&=& \frac{SL(4,\mathbb{R})}{SO(4)}
\;. \label{slsl} \eea
Alternatively, evaluating the Hessian at the solution one finds
 \bea
\nabla\nabla V_{\rm eff} &=&
 4\, z^2 \, P_{1n}\, P_{1n}
\label{eq7_hes} \eea a matrix with $5$ strictly positive eigenvalues
and  $9$ zeros. More precisely the $14$
  scalars in the symmetric traceless $\mathbf{14}$ of $SL\left( 5,\mathbb{R}\right)$ decompose under $SO(4)\sim SU(2)\times SU(2)$
  into the following representations
\be
\mathbf{14}\longrightarrow \underset{m^{2}>0}{\underbrace{\left( \mathbf{2},%
\mathbf{1}\right) \oplus \left( \mathbf{1},\mathbf{2}\right) \oplus
\left( \mathbf{1},\mathbf{1}\right) }}\oplus
\underset{m^{2}=0}{\underbrace{\left( \mathbf{3},\mathbf{3}\right)
}}. \ee
Notice that these $9$ moduli, together with the free radius
$r_1$ and the $10$ degrees of freedom associated to Wilson lines of
the 10 vector fields along $S^1$ sum up to 20 free parameters
characterizing the solution. This precisely matches the dimension of
the moduli space of the six-dimensional solution of which the
present solution is a lift. In other words, in going from six to
seven dimensions, an 11-dimensional
 part of the moduli space translates into
  ``geometrical moduli" describing the circle radius and Wilson lines.
This will be always the case for all $D>6$ solutions under consideration here.

It is worth noticing that the solution with $\mathcal{I}_{2}\neq 0$
can be considered as an intersection of one $\frac{1}{2}$-BPS
electric string and
one $\frac{1}{2}$-BPS magnetic black-$3$-brane, respectively in the $\mathbf{%
5}^{\prime }$ and $\mathbf{5}$ of the $D=7$ $U$-duality group $SL\left( 5,%
\mathbb{R}\right) $ \cite{Ferrara:1997ci}. The corresponding
supporting
charge orbit is $\frac{SL\left( 5,\mathbb{R}\right) }{SL\left( 4,\mathbb{R}%
\right) \times _{s}\mathbb{R}^{4}}$ \cite{Lu:1997bg}, but the common
stabilizer of the two charge vectors is $SL\left(
4,\mathbb{R}\right) $ only, with resulting moduli space of the
considered intersecting configuration given by Eq.~(\ref{slsl}).


\subsection{$AdS_3\times S^2\times T^2$}


We start from the near-horizon ansatz:
\bea
ds^2 &=& r_{\!\rm AdS}^2\, ds^2_{\!\rm AdS_3} + r_{\!S}^2 \, ds^2_{S^2} +
r_1^2 d\theta_1{}^2+
r_2^2 d\theta_2{}^2
\;,
\nonumber\\[1ex]
F_2^{ij} &=& p^{ij}\, \alpha_{S^2}
\;,\qquad
H_{3 i}~=~
e_i\,\beta_{\!\rm AdS_3}
+
\sum_{a=1,2}
p_{i\,a}\,  \alpha_{S^2\times S^1_a}
\;,
\eea
where $T^2=S^1_1\times S^1_2$.
In particular, in this case there are two magnetic three cycles
$S^2\times S^1_a$ which we label by $a=1, 2$.
The corresponding central charges are given by
\bea
Z_{{\rm mag},\,mn} &=&
J_2\, ({\cal V}^{-1})^m{}_i({\cal V}^{-1})^n{}_j\,p^{ij}\;,\nonumber\\[1ex]
Z_{{\rm mag},m}^{\,a}& =& J_{3b} \delta^{ab}\,{\cal V}^j{}_m\,p_{j\,b}\quad
Z_{{\rm el},\,m} = J'_3\, ({\cal V}^{-1})^m{}_i\,q^i
\;,
\label{cc7}
\eea
with $J_2=\left({v_{\rm\!AdS_3}  v_{T^2} \over v_{\!S^2} }\right)^{1\over 2}$,
$J'_3=\left({v_{\rm\!AdS_3}\over v_{\!S^2}  v_{T^2} }\right)^{1\over 2}$,
$J_{3 a}= \left({v_{\rm\!AdS_3} v_{T^2}  \over v_{\!S^2 }\, (v_{S^1_a})^2  }\right)^{1\over 2}$.

 In terms of these charges one can built two U-duality  invariants
 \bea
   {\cal I}_3~=~ \ft18 \epsilon_{ijklm}\, p^{ij} p^{kl} q^m\;, \qquad
   \tilde{\cal I}_3 = \ft12 p_{i\,a}\,p_{j\,b}\,p^{ij}\,\epsilon^{ab} \;.
\eea
Note that the existence of $\tilde{\cal I}_3$ hinges on the fact that there
are two inequivalent magnetic three-cycles.
In terms of the central charges~(\ref{cc7}) the invariants can be written as
 \be
\ft18 \epsilon_{ijklm}\, Z_{\rm mag}^{ij} Z_{\rm mag}^{kl} Z_{\rm el}^m ~=~ J_2^2 J_3'\,{\cal I}_3
\;,
  \qquad
 Z_{{\rm mag} ,i\,a}Z_{{\rm mag} ,j\,b}\,Z_{\rm mag}^{ij}\,\epsilon^{ab} ~=~   J_2 J_{3 a} J_{3 b} \epsilon^{ab} \tilde{\cal I}_3
 \;.
\ee
The effective potential is now given by
\bea
V_{\rm eff}&=&
\ft14Z_{{\rm mag},\,mn} Z_{{\rm mag},\,mn} +
\ft12Z_{{\rm mag}}^{m\,a} Z_{{\rm mag}}^{m\,a} +
\ft12Z_{{\rm el},\,m} Z_{{\rm el},\,m}
\;.
\eea
The Maurer-Cartan equations give $\nabla {\cal V}^i{}_m = {\cal V}^i{}_n P_{mn}$
with $P_{mn}$ symmetric and traceless. The variation of
the central charges is thus given by
\bea
\nabla Z_{{\rm mag},\,mn} &=&  2\, Z_{{\rm mag},\,k [m} P_{n ] k}
\;,\qquad
\nabla Z_{{\rm mag}}^{m\,a}  ~=~  Z_{{\rm mag}}^{n\,a}  P_{mn}
\;,\qquad
\nabla Z_{{\rm el},\,m}  ~=~ -Z_{{\rm el},\,n}  P_{mn}\nn
\;.
\eea
Hence, we obtain
\bea
\nabla V_{\rm eff} &=&
\Big(
-Z_{{\rm mag},\,mk} Z_{{\rm mag},\,nk} +
Z_{{\rm mag}}^{m\,a} Z_{{\rm mag}}^{n\,a} -
Z_{{\rm el},\,m} Z_{{\rm el},\,n}
\Big)\, P_{mn}
~\stackrel{!}\equiv~ 0
\;,
\label{eq7_2}
\eea
By $SO(5)$ rotation, the antisymmetric matrix $Z_{{\rm mag},\,mn}$
can be brought into skew-diagonal form
\bea
Z_{{\rm mag},\,mn} &=&
2z_1 \, \delta_{1 [m} \delta_{n]2}+2z_2 \, \delta_{3 [m} \delta_{n]4}
\;.
\eea
Plugging this into the attractor  equation~(\ref{eq7_2})
 one finds the following solutions
\bea
\mbox{\bf A)}\qquad &&
Z_{{\rm mag},\,mn}= 2z \,( \delta_{1 [m} \delta_{n]2}  +  \delta_{3 [m} \delta_{n]4}    )
 \;,
\quad
Z_{{\rm el},\,m} =z \delta_{m5}
\;,\qquad
Z_{{\rm mag}}^{m\,a} =0
\;.
\qquad\qquad
\label{sol7_2A}
\\[2ex]
\mbox{\bf B)}\qquad &&
 Z_{{\rm mag},\,mn}= 2z \, \delta_{1 [m} \delta_{n]2}  \;,
\qquad
Z_{{\rm el},\,m} =0
\;,
\qquad
Z_{{\rm mag},m}^{a}= z\,\delta_m^{a}
\;.
\label{sol7_2B}
\eea
The corresponding  effective potentials are given by \bea V_{{\rm
eff},A}   &=& \ft32 z^2=\ft32\, (J_2{}^{2}J'_3\,|{\cal I}_3|)^{2/3}
~=~ \frac32 \frac{   v_{\rm\!AdS_3}^3\, v_{T^2}^{1/3} } {v_{\!S^2}\, }
\;|{\cal I}_3|^{2/3}
\;, \nn\\
V_{{\rm eff},B}   &=&
\ft32 z^2=\ft32\, (J_2{}J_{3,1} J_{3,2}\,|\tilde{\cal I}_3|)^{2/3}
~=~
\frac32
\frac{   v_{\rm\!AdS_3}^3\, v_{T^2}^{1/3} }
{v_{\!S^2}\, }
\;|\tilde{\cal I}_3|^{2/3}
\;,
\eea
 in agreement with (\ref{sumads3s2}) with ${\bf I_3}={\cal I}_3$ and  ${\bf I_3}=\tilde{\cal I}_3$ for the solutions ${\bf A}$ and ${\bf B}$, respectively.
After the $\vec{r}$-extremization one finds again that the AdS and sphere radii and the entropy  function
are given by the general formula (\ref{sumads3s2}).
Again, the radii $r_a$ of the two circles $S^1_a$ are not fixed by the
extremization equations.

Let us finally consider the associated moduli spaces. We start with
solution  ${\bf A}$. The symmetry of (\ref{sol7_2A}) is
$Sp(4,\mathbb{R})\sim SO(3,2)$. The moduli space is the quotient of
this group by its maximal compact subgroup \bea {\cal M}_{{\rm
BPS},A}&=& \frac{SO(3,2)}{SO(3)\times SO(2)} \;.
 \label{mbps7A}
\eea
  More precisely, in terms of $SO(3)\times SO(2)$ representations one finds
  that the $14$ scalar components decompose according to
  \be
 P_{mn}: \quad  \mathbf{14}\longrightarrow \underset{m^{2}=0}{\underbrace{\mathbf{3}%
_{+}\oplus \mathbf{3}_{-}}}\oplus \underset{m^{2}>0}{\underbrace{\mathbf{1}%
_{+2}\oplus \mathbf{1}_{-2}\oplus \mathbf{1}_{0}\oplus
\mathbf{5}_{0}}},
  \ee
  subscripts referring to $SO(2)$ charges.
Indeed, evaluating the Hessian
 \bea
\nabla \nabla V_{\rm eff} &=&
2\, P_{mp}P_{nq}\, (Z_{{\rm mag},mp}Z_{{\rm mag},nq}
+Z_{{\rm mag},\,mk} Z_{{\rm mag},\,nk}\,\delta_{pq}  +
Z_{{\rm mag}}^{m\,a} Z_{{\rm mag}}^{n\,a} \,\delta_{pq}\nn\\
&&\qquad\qquad\qquad +
Z_{{\rm el},\,m} Z_{{\rm el},\,n}\,\delta_{pq}
\Big)\;,\nn
\label{eq7_2hes}
\eea
 at the solution (\ref{sol7_2A}) one finds a matrix with
 6 vanishing and 8 strictly positive eigenvalues.

For solution ${\bf B}$ one finds as a symmetry $SL(3,\mathbb{R})$
  and thus the moduli space
 \bea
{\cal M}_{{\rm BPS},B} &=&\frac{SL(3,\mathbb{R})}{SO(3)}\;.
\label{mbps7B} \eea
  The scalars decompose into $SO(3)$ representations according to
   \be
 P_{mn}: \quad  \mathbf{14}\longrightarrow \underset{m^{2}=0}{\mathbf{5}}\oplus \underset{%
m^{2}>0}{\underbrace{2\cdot \mathbf{3}\oplus 3\cdot \mathbf{1}}}.
  \ee
 Indeed from the Hessian
\bea
 \nabla \nabla V_{\rm eff}|_{\bf B} &=& 6 z^2\, (P_{1p}P_{1p}+P_{2p}P_{2p})
 \;,
\eea
 one finds a matrix with 9 strictly positive and 5 vanishing eigenvalues.

 As mentioned above, at $D=7$ the charge orbit supporting one $\frac{1}{2}$%
-BPS black string (or black-$2$-brane) is given by $\frac{SL\left( 5,\mathbb{%
R}\right) }{SL\left( 4,\mathbb{R}\right) \times _{s}\mathbb{R}^{4}}$
\cite {Lu:1997bg}. On the other hand, the charge orbit supporting
one black hole
(or black-$3$-brane) is $\frac{SL\left( 5,\mathbb{R}\right) }{SL\left( 3,%
\mathbb{R}\right) \times SL\left( 2,\mathbb{R}\right) \times _{s}\mathbb{R}%
^{6}}$ and $\frac{SL\left( 5,\mathbb{R}\right) }{SO\left( 3,2\right)
\times _{s}\mathbb{R}^{4}}$ in the $\frac{1}{2}$-BPS and
$\frac{1}{4}$-BPS cases, respectively \cite{Lu:1997bg}.

Solution A corresponds to an intersection of one $\frac{1}{4}$-BPS black-$3$%
-brane (with charges in the $\mathbf{10}^{\prime }$ of
$SL(5,\mathbb{R})$)
with one $\frac{1}{2}$-BPS black string (with charges on the $\mathbf{5}%
^{\prime }$ of $SL(5,\mathbb{R})$). The stabilizer of both charge
vectors is $SO\left( 3,2\right) $ only, and thus the resulting
moduli space of the considered intersecting configuration is given
by Eq.~(\ref{mbps7A}).

Solution B corresponds to an intersection of one $\frac{1}{2}$-BPS black-$3$%
-brane with two parallel $\frac{1}{2}$-BPS black $2$-branes (with
charges on two different $\mathbf{5}$s of $SL(5,\mathbb{R})$).
Accordingly, the stabilizer of the three charge vectors is $SL\left(
3,\mathbb{R}\right) $ only, and thus the resulting moduli space of
the considered intersecting configuration is given by Eq.~(\ref{mbps7B}).


\subsection{$AdS_2\times S^3\times T^2$}


The analysis of the black hole solutions is very similar to the previous one of the
black strings replacing electric with magnetic charges and vice versa.
Now we start from the near-horizon ansatz:
\bea
ds^2 &=& r_{\!\rm AdS}^2 ds^2_{\!\rm AdS_2} + r_{\!S}^2 ds^2_{S^3} +
r_1^2 d\theta_1{}^2+
r_2^2 d\theta_2{}^2
\;,
\nonumber\\[1ex]
F_2^{ij} &=& e^{ij}\, \beta_{\!\rm AdS_2}
\;,\qquad
H_{3 i}~=~
p_i\,\alpha_{S^3}
+
\sum_{r=1,2}
e^r_i\,  \beta_{\!\rm AdS_2\times S^1_r}
\;,
\eea
where $r=1,2$ labels the two inequivalent electric three-cycles
and $T^2=S^1_1\times S^1_2$.
 The central charges and U-duality invariants  are given by
\bea
Z_{{\rm el},mn} &=&
J_2'\, \,{\cal V}^i{}_m{\cal V}^j{}_n\,q_{ij}
\;,\nonumber\\[1ex]
Z_{{\rm el},m}^r &=& J'_{3s}\,\delta^{rs}  \, ({\cal V}^{-1})^m{}_j\,q^j_s
\;,\qquad
Z_{{\rm mag},m} ~=~ J_3\, \,{\cal V}^i{}_m\,p_i \;,\nn\\[1ex]
{\cal I}_3' &=&
\ft18 \epsilon^{ijklm}\, q_{ij} q_{kl} p_m=\ft18 \epsilon^{ijklm}\, Z_{{\rm el},ij} Z_{{\rm el},kl} Z_{\rm mag,m}
\,(J_2^{'2} J_3)^{-1}\;,\nn\\
 \tilde{\cal I}'_3 &=& \ft12 q^{i}_r q^{j}_s\,q_{ij}\,\epsilon^{rs}=\ft12  Z_{{\rm el},r}^i Z_{{\rm el},s}^j
 Z_{{\rm el},ij}  \,\epsilon^{rs} \,(J_2^{'} J'_{3,1} J'_{3,2} )^{-1}
\;,
\eea
with $J_2^{'}=\left({v_{\rm\!AdS_2} \over   v_T  v_{\!S^3} }\right)^{1\over 2}$,
$J_3=\left({  v_T  v_{\rm\!AdS_2} \over  v_{\!S^3} }\right)^{1\over 2}$,
$J'_{3s} = \left({v_{\rm\!AdS_2}  (v_{S^1_s})^2 \over   v_{\!S^3}\, v_{T^2}}\right)^{1\over 2}$.

 The two solutions of the associated attractor equations are given by
\bea
\mbox{\bf A)}\quad &&
Z_{{\rm el},mn}=2\, z (\delta_{1[m}\delta_{n]2}+ \delta_{3[m}\delta_{n]4})
 \;,
\qquad
Z_{{\rm mag},\,m} =z\delta_{m5}
\;,\qquad
Z_{{\rm el},m}^{a} =0
\;.
\label{sol7_2Ap}
\\[1.5ex]
\mbox{\bf B)}\quad &&
Z_{{\rm el},mn}=2\,z \,\delta_{1[m}\delta_{n]2}
 \;,
\qquad
Z_{{\rm mag},\,m} =0
\;,
\qquad
Z_{{\rm el},m}^{a}= z\,\delta^{a}_m
\;.
\label{sol7_2Bp}
\eea
  The effective potentials, entropy  function and near-horizon geometry are given by the general formula
  (\ref{sumads2s3}) with ${\bf I}_3$ given by ${\cal I}_3'$ and  $\tilde{\cal I}_3'$ for the cases ${\bf A}$
  and ${\bf B}$, respectively.
 The analysis of the moduli spaces is identical to that of the $AdS_3$ cases and the
 results are again given by (\ref{mbps7A}) and  (\ref{mbps7B}), respectively.

Solution A has $\widetilde{\mathcal{I}}'_{3}=0$, which comes from $%
q_{ia}q_{jb}\epsilon ^{ab}=0$, meaning that the two $\mathbf{5}$'s
are
reciprocally parallel. On the other hand, solution B has $\mathcal{I}'_{3}=0$%
; this derives from the condition $\epsilon^{ijklm}q_{ij}q_{kl}=0$ for a $%
D=7$ black hole to be $\frac{1}{2}$-BPS \cite{Ferrara:1997ci}.


\section{Maximal $D=8$}\label{Sect8}


\subsection{$\mathcal{N}=2$, $D=8$ Supersymmetry
Algebra}

The maximal $\mathcal{N}=2$, $D=8$ Poincar\'{e} supersymmetry
algebra has
complex chiral supercharges (as in $D=4$) and $\mathcal{R}$-symmetry $%
SU\left( 2\right) \times U\left( 1\right) =Spin(3) \times Spin(2)$.
Its central extension reads as follows (see \textit{e.g.}\
\cite{Strathdee:1986jr,VanProeyen:1999ni,Townsend:1995gp})
\begin{eqnarray}
\left\{ \mathcal{Q}_{\gamma }^{A},\mathcal{Q}_{\delta
}^{B}\right\} &=&C_{\gamma \delta }Z^{\left( AB\right) }+\gamma
_{\gamma \delta }^{\mu \nu }Z_{\mu \nu }^{\left[ AB\right]
}+\gamma _{\gamma \delta }^{\mu \nu \rho
\lambda }Z_{\mu \nu \rho \lambda }^{\left( AB\right) }~(\text{and~h.c.}) \\
\left\{ \mathcal{Q}_{\gamma }^{A},\mathcal{Q}_{\dot{\delta}\mid
B}\right\} &=&\gamma _{\gamma \dot{\delta}\ }^{\mu }Z_{\mu \mid
B}^{A}+\gamma _{\gamma \dot{\delta}\ }^{\mu \nu \rho }Z_{\mu \nu
\rho \mid B}^{A},
\end{eqnarray}
where $A,B=1,2$, so that the supercharges are $SU\left( 2\right)
$-doublets. The trace part of $Z_{\mu \mid B}^{A}$ is the momentum
$P_{\mu }\delta _{B}^{A}$.

Black holes and their magnetic dual (black $4$-brane) central extensions
 $Z^{\left( AB\right) },Z_{\mu \nu \rho \lambda }^{\left( AB\right) } $ sit in the
$\left(
\mathbf{3},\mathbf{2}\right) $ (and $\left( \mathbf{3}^{\prime },\mathbf{2}%
\right) $) of $SU\left( 2\right) \times U\left( 1\right) $, and they
are
embedded in the $\left( \mathbf{3},\mathbf{2}\right) \mathbf{\ }$ of the $U$%
-duality group $SL\left( 3,\mathbb{R}\right) \times SL\left( 2,\mathbb{R}%
\right) $.

On the other hand, dyonic black membrane central extensions
$Z_{\mu \nu }^{\left[ AB\right]
}$
 are in the $\left( \mathbf{1},%
\mathbf{2}\right) $ of $SU\left( 2\right) \times U\left( 1\right) $,
and they are embedded in the $\left( \mathbf{1},\mathbf{2}\right)
\mathbf{\ }$of $SL\left( 3,\mathbb{R}\right) \times SL\left(
2,\mathbb{R}\right) $.

Black strings and their magnetic dual (black $3$-brane)
central extensions  $Z_{\mu \mid
B}^{A},Z_{\mu \nu
\rho \mid B}^{A} $  sit in the $\left( \mathbf{3}%
^{\prime },\mathbf{1}\right) $ (and $\left(
\mathbf{3},\mathbf{1}\right) $)
of $SU\left( 2\right) \times U\left( 1\right) $ (namely in the adjoint of $%
SU\left( 2\right) $, and they do not carry $U(1)$ charge, because
they are real), and they are embedded in the $\left(
\mathbf{3},\mathbf{1}\right) \mathbf{\ }$ of the $U$-duality group
$SL\left( 3,\mathbb{R}\right) \times SL\left( 2,\mathbb{R}\right) $.

In our analysis, the corresponding central charges are denoted by $%
Z_{{\rm el},iA }$ and $Z_{{\rm mag},iA }$ ($i=1,2,3$ and $A =1,2$) for black
holes and their magnetic duals, by $Z_{{\rm el},i}$ and $Z_{{\rm mag},i}$ for
black strings and their magnetic duals, and by $Z_{A }$ for dyonic
black $2$-branes.

\subsection{$\mathcal{N}=2$, $D=8$ Supergravity}

The bosonic field content of $D=8$ supergravity~\cite{Salam:1984ft}
with maximal supersymmetry
includes beside the graviton, scalars in the symmetric manifold
 \bea
{\cal V}_i{}^m, {\cal V}_A{}^B: && { SL(3,\mathbb{R})\over SO(3)}
\times { SL(2,\mathbb{R})\over SO(2)} \;, \eea with $i,m=1, \dots,
3$, $A=1,2$, and forms in the following representations of the
$SL(3,\mathbb{R})\times SL(2,\mathbb{R})$ U-duality group: \bea
  F_2^{iA}: &&       ({\bf 3},{\bf 2} ) \;,\nn\\
  H_{3 i}   : &&      ({\bf 3'},{\bf 1})\;,\nn\\
  F_4^{A} : && ({\bf 1},{\bf 2}) \;.
 \eea
The Lagrangian carries only one three-form potential $C$, whose
field strength $F_4$ together with its magnetic dual spans the
$SL(2,\mathbb{R})$ doublet $F_4^{A}$.

The scalar vielbeins ${\cal V}_i{}^m$, ${\cal V}_A{}^B$,
corresponding to the two factors, vary as \bea \nabla {\cal
V}_i{}^m &=& {\cal V}_i{}^n P_{mn}\;,\qquad \nabla {\cal V}_A{}^B
~=~ {\cal V}_A{}^C\,P_{BC} \;, \label{MC_8} \eea with $P_{mn}$ and
$P_{AB}$, symmetric and traceless. Here we raise and lower indices
$m$ and $A$ with $\delta_{mn}$ and $\delta_{BA}$, respectively.

\


\subsection{$AdS_3 \times S^3 \times T^2$}


We start with the $AdS_3 \times S^3 \times T^2$ near-horizon
ansatz: \bea ds^2 &=& r_{\!\rm AdS}^2\, ds^2_{\!\rm AdS_3} +
r_{\!S}^2 \, ds^2_{S^3} + \sum_{s=1,2} r_s^2 d\theta_s{}^2 \;,
\qquad F_{2,iA }=q_{iA }\alpha_{T^{2}},
\nonumber\\[1ex]
H_{3 i}&=& p_i\,\alpha_{S^3} +e_i\,\beta_{\!\rm AdS_3} \;,\qquad
F_{4} ~=~ \sum_{r=1,2} e_r\,\beta_{\!\rm AdS_3\times S^1_r}
+\sum_{a=1,2} p_a\,\alpha_{S_3\times S^1_a} \;. \eea Depending on
the choice of the circle within $T^2=S^1_1\times S^1_2$, there are
two inequivalent electric and magnetic four-cycles. The four-form
charges combine into the $SL(2,\mathbb{R})$ doublet $Q_{A\,r}=(q_r,
p_r)$.

The central charges are given by
\bea
Z_{{\rm mag},m} &=&
J_3\,({\cal V}^{-1})_m{}^j\,p_j
\;,\quad
Z_{{\rm el}}^m ~=~
J_3'\,{\cal V}_i{}^m\,q^i
\;,
\quad Z^r_{A} = J_{4s} \delta^{sr}\,({\cal
V}^{-1})_A{}^B\,Q_{B\,s} \;, \eea with $J_3=\left(\frac{
v_{\rm\!AdS_3} \,v_{T^2} }{v_{\!S}^3}\right)^{\frac12}$,
$J'_3=\left(\frac{v_{\rm\!AdS_3}}{ v_{\!S^3}\,v_{T^2}
}\right)^{\frac12}$, $J_{4s}= \left(\frac{v_{\rm\!AdS_3}\,v_{T^2}}{
v_{\!S^3}\,(v_{S^1_s})^2}\right)^{\frac12}$.

The U-duality invariants that can be built with these set of charges are
\bea
{\cal I}_2 &=& p_i\,q^i = Z_{{\rm el},i} Z_{{\rm mag},i} (J_3 J_3')^{-1}
\;,
\nn\\[.5ex]
 \tilde{\cal I}_2 &=&
q_{A\,r}q_{B\,s}\,\epsilon^{AB}\epsilon^{rs}= Z_{{\rm
el},A\,r}Z_{{\rm el}B\,s}\,\epsilon^{AB}\epsilon^{rs}
(J_{4,1}J_{4,2} )^{-1} \;. \eea Note that the existence of two
inequivalent electric four-cycles is crucial for the existence of
$\tilde{\cal I}_2$. The effective potential can be written as \bea
V_{\rm eff} &=& \ft12 Z_{{\rm mag},m}Z_{{\rm mag},m} +\ft 12
Z_{{\rm el}}^m Z_{{\rm el}}^m + \ft12 Z_{A\,r}Z_{A\,r} \;, \eea
and the attractor equations take the form \bea \Big(Z_{{\rm
mag},m}Z_{{\rm mag},n}- Z_{{\rm el}}^m Z_{{\rm el}}^n\Big)\,P_{mn}
+ \Big(Z_{A\,r}Z_{B\,r}\Big)\,P_{AB} &\stackrel1{\equiv}& 0 \;.
\eea We will consider the following two solutions to these
equations \bea \mbox{\bf A)}\qquad && Z_{{\rm mag},m} =\pm Z_{{\rm
el}}^m=z \,\delta_{m1} \;,\qquad Z_{A\,r}=0 \label{sol81}
\;.\\[1ex]
\mbox{\bf B)}\qquad && Z_{{\rm mag},m} =Z_{{\rm el}}^m=0\;,\qquad
Z_{A\,r}=z\,\delta_{A r} \;. \label{sol82} \eea The effective
potentials at the horizon become
\bea
V_{\rm eff} |_{\bf A} &=& z^2=
J_3J_3'\,|{\cal I}_2| ~=~ \frac{v_{\rm\!AdS_3}}{ v_{\!S^3}}\,|{\cal I}_2|\;,\nn\\
V_{\rm eff} |_{\bf B} &=& z^2=
J_{4,1}J_{4,2}\,|\tilde{\cal I}_2| ~=~ \frac{ v_{\rm\!AdS_3}}{ v_{\!S^3}}\,|\tilde{\cal I}_2|
\;,
\eea
respectively. Plugging this into the entropy  function and extremizing
with respect to the radii one recovers the near-horizon geometry
central charge (\ref{sumads3s3}) with ${\bf I}_2$ taken as ${\cal I}_2$ or $\tilde{\cal I}_2$ for the solution ${\bf A}$ and ${\bf B}$, respectively.

Let us consider the associated moduli spaces. The symmetry groups
leaving (\ref{sol81}) and (\ref{sol82}) invariant are
$SL(2,\mathbb{R})^2$ and $SL(3,\mathbb{R})$, respectively. The
moduli spaces are thus \bea {\cal M}_{{\rm BPS},A} &=&  \left(
{SL(2,\mathbb{R})   \over SO(2)} \right)^2 \;,\qquad\qquad {\cal
M}_{{\rm BPS},B} ~=~  {SL(3,\mathbb{R}) \over SO(3)}
\;.\label{modulid8ads3s3}
 \eea
 The same results  follow from evaluating the Hessians at the solutions
 \bea
 \nabla \nabla V_{{\rm eff},A} &=& 2 z^2 P_{1m}^2\;,\qquad\qquad
  \nabla \nabla V_{{\rm eff},B} ~=~ 2 z^2 P_{A A}^2
\;,
\eea
which shows that one has 3(2) strictly positive and 4(5) vanishing eigenvalues  for the
solution ${\bf A}({\bf B})$, in agreement with the dimensions  of the moduli spaces (\ref{modulid8ads3s3}).

At $D=8$ there are two dyonic $\frac{1}{2}$-BPS black-$2$-brane,
whose charge
orbits are $\frac{SL\left( 3,\mathbb{R}\right) \times SL\left( 2,\mathbb{R}%
\right) }{SL\left( 3,\mathbb{R}\right) \times \mathbb{R}}$. The $\frac{1}{2}$%
-BPS black strings (and their dual black $3$-branes) are in the
$\left(
\mathbf{3}^{\prime },\mathbf{1}\right) $ and $\left( \mathbf{3},\mathbf{1}%
\right) $ of the $D=8$ $U$-duality group $SL\left(
3,\mathbb{R}\right)
\times SL\left( 2,\mathbb{R}\right) $, and their individual charge orbit is $%
\frac{SL\left( 3,\mathbb{R}\right) \times SL\left( 2,\mathbb{R}\right) }{%
\left( SL\left( 2,\mathbb{R}\right) \times _{s}\mathbb{R}^{2}\right)
\times SL\left( 2,\mathbb{R}\right) }$ \cite{Lu:1997bg}. The black
holes (and their dual black $4$-branes) are in the $\left(
\mathbf{3},\mathbf{2}\right) $ and
$\left( \mathbf{3}^{\prime },\mathbf{2}\right) $ of $SL\left( 3,\mathbb{R}%
\right) \times SL\left( 2,\mathbb{R}\right) $, and their individual
charge
orbit is $\frac{SL\left( 3,\mathbb{R}\right) \times SL\left( 2,\mathbb{R}%
\right) }{SL\left( 2,\mathbb{R}\right) \times _{s}\mathbb{R}^{2}}$ and $%
\frac{SL\left( 3,\mathbb{R}\right) \times SL\left( 2,\mathbb{R}\right) }{%
GL\left( 2,\mathbb{R}\right) \times _{s}\mathbb{R}^{3}}$ for the
$\frac{1}{4} $-BPS and $\frac{1}{2}$-BPS cases, respectively
\cite{Lu:1997bg}.

In the considered $AdS_{3}\times S^{3}\times T^{2}$ near-horizon
geometry, solution A corresponds to the intersection of
one $\frac{1}{2}$-BPS black string and one $\frac{1}{2}$-BPS black
$3$-brane. The stabilizer of both
charge vectors is $SL\left( 2,\mathbb{R}\right) \times SL\left( 2,\mathbb{R}%
\right) $ only, and thus the resulting moduli space of the
considered
interesecting configuration is $\left( \frac{SL\left( 2,\mathbb{R}\right) }{%
SO\left( 2\right) }\right) ^{2}$, given in the left hand side of Eq.~(\ref {modulid8ads3s3}).

On the other hand, solution B corresponds to the instersection of
two dyonic $\frac{1}{2}$-BPS black $2$-branes. Thus the stabilizer
of both charge vectors is $SL\left( 3,\mathbb{R}\right) $ only, and
the resulting moduli
space of the considered interesecting configuration is $\frac{SL\left( 3,%
\mathbb{R}\right) }{SO\left( 3\right) }$, given in the right-hand
side of Eq.~(\ref{modulid8ads3s3}).


\subsection{$AdS_3 \times S^2 \times T^3$}


The near-horizon ansatz is given by
\bea
ds^2 &=& r_{\!\rm AdS}^2\, ds^2_{\!\rm AdS_3} + r_{\!S}^2 \, ds^2_{S^2} +
\sum_{s=1,2,3} r_s^2 d\theta_s{}^2
\;,
\nonumber\\[1ex]
F_{2}^{iA} &=& p^{iA}\,\alpha_{S^2} \;,\qquad H_{3 i}~=~
\sum_{a=1,2,3}\,p^a_i\,\alpha_{S^2\times S^1_a}
\;,\nonumber\\[1ex]
F_{4}^A &=& \sum_{a=1,2,3} e^A_a\,\beta_{\!\rm AdS_3\times S^1_a}
+\sum_{a=1,2,3}\ft12\,|\epsilon_{abc}|\,
 p^A_a\,\alpha_{(S^2\times S^1_b \times S^1_c)} \;,
\eea with $T^3=S^1_1\times S^1_2\times S^1_3$. In this near-horizon
geometry there are thus three inequivalent electric and magnetic
four-cycles and three magnetic three-cycles. The four-form charges
again combine into the $SL(2,\mathbb{R})$ doublet $Q_{A\,a}=(q_a,
p_a)$.

The central charges take the form \bea Z_{{\rm mag}}^{mA} &=&
J_2\, {\cal V}_B{}^A\, {\cal V}_k{}^m\, p^{kB} \;, \qquad
Z^a_{{\rm mag},m} ~=~ J_{3 b} \delta^{ba}\,({\cal
V}^{-1})_m{}^j\,p_j^b \;,
\nonumber\\[.5ex]
Z_{A a} &=& J_{4\,b} \delta_{ab}\,({\cal V}^{-1})_A{}^B\,Q_{B\,b}
\;, \eea with $J_2=\left(\frac{v_{\rm\!AdS_3} \,v_{T^3}  }{ v_{\!S^2}
}\right)^{\frac12}$, $J_{3a}=\left(\frac{ v_{\rm\!AdS_3}  \,v_{T^3}
}{ v_{\!S^2}\,(v_{S^1_a})^2 }\right)^{\frac12}$, $J_{4
a}=\left(\frac{ v_{\rm\!AdS_3} \,(v_{S^1_a})^2 }{v_{\!S^2}\,v_{T^3}
}\right)^{\frac12}$.

The  U-duality invariants are given by
\bea
 {\cal I}_3 &=& p^{iA}\,p^a_i\,Q_{A a} =Z_{\rm mag}^{iA}\,Z_{{\rm mag},i a}\,Z_{A a}
 \,(J_2  J_{3a} J_{4a} )^{-1}\;,\nn\\
\tilde{\cal I}_3 &=&
\ft13 p^a_ip^b_jp^c_k \,\epsilon^{ijk}\epsilon_{abc}=\ft13
Z_{{\rm mag},i}^a Z_{{\rm mag},j}^b   Z_{{\rm mag},k}^c
 \,\epsilon^{ijk}\epsilon_{abc}\, (J_{3,1}  J_{3,2} J_{3,3} )^{-1}
\;.
\eea
 From variation of the effective potential
\bea V_{\rm eff}&=& \ft12Z_{{\rm mag}}^{mA}Z_{{\rm mag}}^{mA}+
\ft12Z^a_{{\rm mag},m}Z^a_{{\rm mag},m}+ \ft12Z_{A\,r} Z_{A\,r}
\;, \eea we thus obtain the attractor equations \bea \Big(Z_{{\rm
mag}}^{mA}Z_{{\rm mag}}^{nA}- Z^a_{{\rm mag}\,m}Z^a_{{\rm
mag}\,n}\Big)\,P_{mn} &\stackrel{!}\equiv& 0 \;,
\nonumber\\[.5ex]
\Big(Z_{{\rm mag}}^{mA}Z_{{\rm mag}}^{mB} -Z_{A\,r} Z_{B\,r}
\Big)\,P_{AB} &\stackrel{!}\equiv& 0 \;, \label{eq8} \eea with
$P_{mn}$ and $P_{AB}$ symmetric and traceless.

We will consider the solutions \bea \mbox{\bf A)}\qquad && Z_{{\rm
mag}}^{mA}~=~0~=~Z_{A\,a}\;,\qquad Z^a_{{\rm mag},m}~=~
z\,\delta^a_m
\;.\label{sol8A}\\[1ex]
\mbox{\bf B)}\qquad && Z_{{\rm mag}}^{mA} = z\,
\delta_{m1}\delta_{A 1}\;, \quad Z^a_{{\rm
mag},m}=\delta^{a1}\delta_{m1}\,z\;,\quad Z_{A\,r} =\delta_{r1}\,
\delta_{A 1}\,z \;. \label{sol8} \eea The effective potentials
become \bea V_{{\rm eff},A}   &=& \ft32 z^2=\ft32\, (J_2 J_{3a}
J_{4a} \,|\tilde{\cal I}_3)^{2/3} ~=~ \frac{ 3\, v_{\rm\!AdS_3}\,
v_{T^3}^{1\over 3}  } {2 v_{\!S^2} } \;|\tilde{\cal I}_3|^{2/3}
\;, \nn\\
V_{{\rm eff},B}   &=& \ft32 z^2=\ft32\, (J_{3,1}J_{3,2}
J_{3,3}\,|{\cal I}_3|)^{2/3} ~=~ \frac{  3\, v_{\rm\!AdS_3}\,
v_{T^3}^{1\over 3}  } {2 v_{\!S^2} } \;|{\cal I}_3|^{2/3} \;, \eea
respectively, and $\vec{r}$-extremization leads to the near-horizon
geometry and central charge (\ref{sumads3s2}) with ${\bf
I}_3=\tilde{\cal I}_3$   and ${\bf I}_3={\cal I}_3$ for the cases
 {\bf A} and {\bf B}, respectively.

The symmetry group leaving (\ref{sol8A}) and (\ref{sol8}) invariant,
is $SL(2,\mathbb{R})$ and the moduli space thus given by \bea {\cal
M}_{{\rm BPS},A/B} &=&  {SL(2,\mathbb{R})   \over SO(2)} \;.
 \label{modulid8ads3s2}
 \eea
Alternatively the moduli space can be determined
from the vanishing eigenvalues of the Hessians
 \bea
 \nabla \nabla V_{{\rm eff},A} &=& 2 z^2 (P_{1m}^2+P_{2m}^2)\;,\nn\\
  \nabla \nabla V_{{\rm eff},B} &=& 4 z^2 P_{1n}^2+6 z^2 P_{1 A}^2\;,
 \eea
respectively,
showing 5 strictly positive and 2 vanishing eigenvalues in each case.

We now derive the nature of the moduli spaces of solutions A and B
from the charge orbits discussed in \cite{Lu:1997bg}.

Solution A corresponds to an intersection of three black $3$-branes, with $%
\widetilde{\mathcal{I}}_{3}=\det\left( p_{i}^{a}\right) \neq 0$ but $\mathcal{%
I}_{3}=p^{iA }p_{i}^{a}Q_{A a}=0$. The charge orbit for each of
them is $\frac{SL\left( 3,\mathbb{R}\right) \times SL\left( 2,\mathbb{R}%
\right) }{\left( SL\left( 2,\mathbb{R}\right) \times
\mathbb{R}^{2}\right)
\times SL\left( 2,\mathbb{R}\right) }$, and the common stabilizer is the $%
SL\left( 2,\mathbb{R}\right) $ commuting with $SL\left( 3,\mathbb{R}\right) $%
. This agrees with the moduli space $\left. \frac{SL\left( 2,\mathbb{R}%
\right) }{SO\left( 2\right) }\right| _{A}$ of solution A (see Eq.~(\ref{modulid8ads3s2})).

Solution B corresponds to the intersection of three parallel black $2$%
-branes, three parallel black $3$-branes and $\frac{1}{2}$-BPS black $4$%
-branes, respectively characterized by the constraints
\begin{equation}
Q_{A a}Q_{B b}\epsilon ^{A B }=0,~~p_{i}^{a}p_{j}^{b}\epsilon
^{ijk}=0,~~~p^{iA }p^{jB }\epsilon _{A B }=0,
\end{equation}
with $\mathcal{I}_{3}=p^{iA }p_{i}^{a}Q_{A a}\neq 0$ and $%
\widetilde{\mathcal{I}}_{3}=0$.

The three parallel $3$-branes have a common charge orbit $\frac{SL\left( 3,%
\mathbb{R}\right) \times SL\left( 2,\mathbb{R}\right) }{\left( SL\left( 2,%
\mathbb{R}\right) \times _{s}\mathbb{R}^{2}\right) \times SL\left( 2,\mathbb{%
R}\right) }$, whereas the parallel $2$-branes have a common charge orbit $%
\frac{SL\left( 3,\mathbb{R}\right) \times SL\left( 2,\mathbb{R}\right) }{%
SL\left( 3,\mathbb{R}\right) \times \mathbb{R}^{1}}$, and the $\frac{1}{2}$%
-BPS $4$-brane has charge orbit\ $\frac{SL\left( 3,\mathbb{R}\right)
\times SL\left( 2,\mathbb{R}\right) }{\left( GL\left(
3,\mathbb{R}\right) \times _{s}\mathbb{R}^{2}\right) \times
\mathbb{R}^{1}}$ \cite{Lu:1997bg}.

Since the coset is factorized, the common stabilizer of the three
parallel $3 $-branes and of $\frac{1}{2}$-BPS $4$-brane is $SL\left(
2,\mathbb{R}\right) $ inside $SL\left( 3,\mathbb{R}\right) $, and
this agrees with the moduli
space $\left. \frac{SL\left( 2,\mathbb{R}\right) }{SO\left( 2\right) }%
\right| _{B}$ of solution B (see Eq.~(\ref{modulid8ads3s2})).


\subsection{$AdS_2 \times S^3 \times T^3$}


This case is very similar to the previous discussion.
We start with the near-horizon ansatz
\bea
ds^2 &=& r_{\!\rm AdS}^2\, ds^2_{\!\rm AdS_2} + r_{\!S}^2 \, ds^2_{S^3} +
\sum_{s=1,2,3} r_s^2 d\theta_s{}^2
\;,
\nonumber\\[1ex]
F_{2}^{iA} &=& e^{iA}\,\beta_{\!\rm AdS_2} \;,\qquad H_{3 i}~=~
\sum_{a=1,2,3}\,e^r_i\,\beta_{\!\rm AdS_2\times S^1_r}
\;,\nonumber\\[1ex]
F_{4} &=& \sum_{a,b,c=1,2,3}\ft12 |\epsilon_{abc}| \,
e^A_c\,\beta_{\!\rm AdS_2\times S^1_a\times S^1_b} +\sum_{a=1,2,3}
p^A_a\,\alpha_{S_3\times S^1_a} \;, \eea with $T^3=S^1_1\times
S^1_2\times S^1_3$. Again, there are thus three inequivalent
electric and magnetic four-cycles. In addition, there are three
inequivalent electric three-cycles.

The associated central charges and U-duality invariants are given
by \bea Z_{{\rm el},mA} &=& J'_2\, ({\cal V}^{-1})_A{}^B\, ({\cal
V}^{-1})_m{}^k\, q_{kB} \;, \qquad Z^{m\,r}_{{\rm el}} ~=~ J'_{3s}
\delta_{sr}\,{\cal V}_j{}^m\,q^j_s \;,
\nonumber\\[.5ex]
Z_{A\,r} &=& J_{4\,s} \delta_{rs} \,({\cal
V}^{-1})_A{}^B\,Q_{B\,s} \;,
\nonumber\\[1ex]
 {\cal I}_3' &=& q_{iA}\,q^i_a \,Q^{A a} =Z_{{\rm el},iA}\,Z_{\rm el}^{i a}\,Z^A_a
 \,(J_2'  J_{3a}' J_{4a}' )^{-1}\;,\nn\\[.5ex]
\tilde{\cal I}_3' &=&
\ft13 q_a^i q_b^j q_c^k \,\epsilon_{ijk}\epsilon^{abc}=\ft13
Z_{{\rm el},a}^i Z_{{\rm el},b}^j   Z_{{\rm el},c}^k
 \,\epsilon_{ijk}\epsilon^{abc}\, (J_{3,1}  J_{3,2} J_{3,3} )^{-1}
 \;,
\eea
with
$J'_2=\left(\frac{ v_{\rm\!AdS_2} }{v_{\!S}^3\,\, v_{T^3}  }\right)^{\frac12}$,
$J'_{3r}=\left(\frac{v_{\rm\!AdS_2} \,(v_{S^1_r})^2}{v_{\!S^3}\,v_{T^3}   }\right)^{\frac12}$,
$J_{4\,r}=\left(\frac{ v_{\rm\!AdS_2} \,v_{T^3}}{v_{\!S^3}\,(v_{S_r^1})^2 }\right)^{\frac12}$.

The possible solutions of the attractor equations are \bea
\mbox{\bf A)}\qquad && Z_{{\rm el},mA}~=~0~=~Z_{A\,s}\;,\qquad
Z^{m\,r}_{{\rm el}}~=~ z\,\delta^{mr} \;,
\nonumber\\[.5ex]
\mbox{\bf B)}\qquad && Z_{{\rm el},mA}=z\, \delta_{m1}\delta_{A 1}
 \qquad
Z^{m\,r}_{{\rm el}}=\delta^{r1}\delta_{m1}\,z\;,\quad Z_{A\,r}
=\delta_{r1}\, \delta_{A 2}\,z \;. \eea
  The effective potentials, entropy  function and near-horizon geometry are given by the general formula
  (\ref{sumads2s3}) with ${\bf I}_3={\cal I}_3$   and ${\bf I}_3=\tilde{\cal I}_3$ for the cases {\bf A} and {\bf B} respectively.
 The analysis of the moduli spaces is identical to that on $AdS_3\times S^2$ case and the results are given again by
(\ref{modulid8ads3s2}).


\section{The Lift to Eleven Dimensions}\label{Sect9}


The attractor solutions we have discussed
throughout this paper have a simple lift to eleven-dimensional supergravity.
 The black string  solutions  with $AdS_3\times S^3\times T^{D-6}$
 near-horizon geometry follow from
 dimensional reduction  of  M2M5 branes intersecting
 on a string.
   The supersymmetric solutions with $AdS_3\times S^2\times T^{D-5}$ follow from
 reductions of triple M2-intersection on a string. Finally
   $AdS_2\times S^3\times T^{D-5}$ near-horizon geometries correspond to triple
M5 intersections on a time-like line. The orientations of the M2,M5
branes in the three cases are summarized in table~\ref{M-int}.

\begin{table}[h]
\begin{center}
\begin{tabular}{|c|ccccccccccc|c|}
\hline
    & 0 &1&2&3&4&5&6&7&8&9&10& near-horizon\\
    \hline
M2 & $-$ & $-$  & $ \bullet$ &   $ \bullet$ & $ \bullet$ & $ \bullet$ & $ \bullet$ & $ \bullet$ & $ \bullet$ & $ \bullet$ & $ -$ & $AdS_3\times S^3\times T^5$ \\
M5 & $-$ & $-$  & $ \bullet$ &   $ \bullet$ & $ \bullet$ & $ \bullet$ & $ -$ & $ -$ & $ -$ & $ -$ & $ \bullet$ & \\
\hline
M2 & $-$ & $\bullet$  & $ \bullet$ &   $ \bullet$ & $ \bullet$ & $ \bullet$ & $ \bullet$ & $ \bullet$ & $ \bullet$ & $ -$ & $ -$&
 $AdS_2\times S^3\times T^6$  \\
M2 & $-$ & $\bullet$  & $ \bullet$ &   $ \bullet$ & $ \bullet$ & $ \bullet$ & $ \bullet$  & $ -$ & $ -$ & $ \bullet$ & $ \bullet$ & \\
M2 & $-$ & $\bullet$  & $ \bullet$ &   $ \bullet$ & $ \bullet$  & $ -$ & $ -$& $ \bullet$ & $ \bullet$  & $ \bullet$ & $ \bullet$ & \\
\hline
M5 & $-$ & $- $  & $ \bullet$ &   $ \bullet$ & $ \bullet$ & $ \bullet$ & $ \bullet$ & $ -$ & $ -$ & $ -$ & $ -$&
 $AdS_3\times S^2\times T^6$  \\
 M5 & $-$ & $- $  & $ \bullet$ &   $ \bullet$ & $ \bullet$  & $ -$ & $ -$& $ \bullet$ & $ \bullet$ & $ -$ & $ -$&
   \\
 M5 & $-$ & $- $  & $ \bullet$ &   $ \bullet$ & $ \bullet$  & $ -$ & $ -$ & $ -$ & $ -$& $ \bullet$ & $ \bullet$&
   \\
 \hline
\end{tabular}
\end{center}
\caption{\small Supersymmetric M-intersections}
\label{M-int}
\end{table}%

After dimensonal reductions down to $D=6, 7, 8$-dimensions the
solutions expose a variety of charges with respect to forms of
various rank. Indeed, a single brane intersection in $D=11$ leads
to different solutions after reduction to $D$-dimensions depending
on the orientation of the M-branes along the internal space.
Different solutions carry charges with respect to a different set
of forms in the $D$-dimensional supergravity. They can be fully
characterized by U-duality invariants built out of the brane
charges. The list of U-duality invariants leading to extremal
black $p$-brane solutions in $D=6, 7, 8$ dimensions are listed in
table~\ref{chint}. The reader can easily check that there is a
one-to-one correspondence between the entries in this table and
the solutions found in the previous sections.

\begin{table}[h]

\begin{center}
\begin{tabular}{|c| c|c|}
\hline
dim & $p=0$   & $p=1$ \\
\hline
5 &  $q_2 q_2 q_2$   & $p_2 p_2 p_2$ \\
6 &  $q_2 q_2 q_3$   & $p_3 q_3 $, $p_2 p_2 p_3  $   \\
7&  $q_2 q_2 p_3$,   $q_3 q_3 q_2$   & $p_3 q_3 $, $p_2 p_2 q_3 $, $p_3 p_3 p_2$   \\
8&  $q_2 q_3 p_4$,   $q_3 q_3 q_3$   & $p_3 q_3$, $p_4 q_4 $, $p_2 p_3 q_4 $,  $p_3 p_3 p_3$  \\
\hline
\end{tabular}
\end{center}
\caption{\small Electric and magnetic charges for M-brane intersections. $p=0,1$
corresponds to intersections on a black hole and a black string respectively.
$q_n$($p_n$) denotes the electric(magnetic) charge  of the brane solution
and $n$ specifies the rank of the form.}
\label{chint}
\end{table}%

\section{Final Remarks}\label{Sect10}

In the present paper we analyzed the attractor nature of solutions
of some supergravity theories in $D=6$, $7$, $8$, with static,
asymptotically flat, spherically symmetric extremal black-$p$
brane backgrounds and scalar fields turned on.

We have found that for such theories, with the near-horizon
geometry containing a factor $AdS_{p+2}$ ($p=0,1$), a
generalization of the ``entropy function'' \cite{Sen:2005wa} and
``effective potential'' \cite {Ferrara:1995ih,Strom,FK1,FK2,FGK}
formalisms occurs, which allows one to determine the scalar flow
and the related moduli space near the horizon. The value of the
entropy function at its minimum is given in terms of U-duality
invariants built out of the brane charges and it measures the
central charges of the dual CFT living on the AdS boundary. The
resulting central charges were shown to satisfy a
Bekenstein-Hawking like area law generalizing the familiar results
of black hole physics.

In order to make further contact with previous work on $p$-brane
intersections and their supersymmetry-preserving features
\cite{Lu:1997hb}, we have found that for maximal supergravities in
$D$ space-time dimensions, the moduli spaces of attractors with
$AdS_{3}\times S^{3}\times T^{D-6}$
near-horizon geometries have rank $10-D$. Actually, this holds also for the $%
D=4$ case (with near-horizon geometry $AdS_{2}\times S^{2}$) in the
non-BPS configuration, with the related moduli space given by the
rank-$6$ symmetric space $\frac{E_{6\left( 6\right) }}{USp\left(
8\right) }$ \cite{Ferrara-Marrani-2}.

Furthermore, for $D$-dimensional maximal supergravities, the moduli
spaces of attractors with $AdS_{3}\times S^{2}\times T^{D-5}$ (or
$AdS_{2}\times S^{3}\times T^{D-5}$) near-horizon geometries have
rank $9-D$.
This holds also for the $D=5$ case (with near-horizon geometry $%
AdS_{3}\times S^{2}$ or $AdS_{2}\times S^{3}$) in the
$\frac{1}{8}$-BPS configuration, with the related moduli space given
by the rank-$4$ symmetric space $\frac{F_{4\left( 4\right)
}}{USp\left( 2\right) \times USp\left( 6\right) }$
\cite{Andrianopoli:1997hb, Ferrara-Marrani-2}.

These results imply that the dilatons of the $p$-brane intersections in $%
D=11 $ described in \cite{Lu:1997hb} are not all on equal footing,
because only one or two (combinations) of them get(s) fixed at the
horizon, while the other ones have asymptotical values which enter
the flow, although the function $F$ does not depend on such values.

Finally, we would like to comment on the fact that the half-maximal
non-chiral $\left( 1,1\right) $, $D=6$ theory analyzed in Sect.~\ref{Sect5}
may be considered as Type IIA compactified on $K3$
\cite{Seiberg}. The result obtained in the present paper for the
$AdS_{3}\times S^{3}$ near-horizon geometry supports the conjecture
of \cite{Mo-various}. On the other hand, we do not find an agreement
with the other \textit{Ans\"{a}tze} for the near-horizon geometry,
because we only find solutions where the charges of strings (or
that of their magnetic duals) are turned on.

 We note that the techniques we have developed here apply to any supergravity flow
 ending on an AdS horizon even in presence of higher derivative terms and gaugings.
 It would be interesting to apply this formalism to the study of higher derivative
 corrections  to central charges in ungauged and gauged
  supergravities  extending the black holes results
  found in \cite{Chandrasekhar:2006kx} and \cite{Morales:2006gm}.
  The study of  non-BPS black p-brane flows along the lines of \cite{tutti}
 deserves also further investigations.

\section*{\textbf{Acknowledgments}}

We would like to thank M. Bianchi, E. G. Gimon and A. Yeranyan for
interesting discussions.

The work of S. F.~has been supported in part by D.O.E.~grant
DE-FG03-91ER40662, Task C, and by the Miller Institute for Basic
Research in Science, University of California, Berkeley, CA, USA.

A. M. would like to thank the Berkeley Center for Theoretical
Physics (CTP) of the University of California, Berkeley, USA, where
part of this work was done, for kind hospitality and stimulating
environment.
The work of A. M. has been supported by an INFN visiting
Theoretical Fellowship at SITP, Stanford University, Stanford, CA,
USA.

The work of H. S. has been supported in part by the Agence
Nationale de la Recherche (ANR).



\begin{thebibliography}{9}

\bibitem{Salam:1989fm}
  A.~Salam and E.~Sezgin,
  ``\textit{Supergravities in Diverse Dimensions. Vol. 1, 2},''
{\it  Amsterdam, Netherlands: North-Holland (1989) 1499 p.
Singapore, Singapore: World Scientific (1989)}

\bibitem{ADF-U-duality}  L. Andrianopoli, R. D'Auria and S. Ferrara, $%
\mathit{U}$\textit{\ duality and central charges in various
dimensions
revisited}, Int. J. Mod.Phys. \textbf{A13}, 431 (1998), \texttt{%
hep-th/9612105}.

\bibitem{Lu:1997hb}
  H.~Lu, C.~N.~Pope, T.~A.~Tran and K.~W.~Xu,
  \textit{Classification of p-branes, NUTs, waves and intersections},
  Nucl.\ Phys.\   {\bf B511}, 98 (1998), \texttt{hep-th/9708055}.

\bibitem{Ferrara:1995ih}
  S.~Ferrara, R.~Kallosh and A.~Strominger,
  {\it $\mathcal{N}=2$ extremal black holes,},
  Phys.\ Rev.\ {\bf D52}, 5412 (1995), \texttt{hep-th/9508072}.

\bibitem{Strom}  A. Strominger, \textit{Macroscopic Entropy of }$\mathcal{N}%
\mathit{=2}$\textit{\ Extremal Black Holes}, Phys. Lett.
\textbf{B383}, 39 (1996), \texttt{hep-th/9602111}.

\bibitem{FK1}  S. Ferrara and R. Kallosh, \textit{\ Supersymmetry and
Attractors}, Phys. Rev. \textbf{D54}, 1514 (1996),
\texttt{hep-th/9602136}.

\bibitem{FK2}  S. Ferrara and R. Kallosh, \textit{\ Universality of
Supersymmetric Attractors}, Phys. Rev. \textbf{D54}, 1525 (1996), \texttt{%
hep-th/9603090}.

\bibitem{FGK}  S.~Ferrara, G. W. Gibbons and R. Kallosh, \textit{Black Holes
and Critical Points in Moduli Space}, Nucl. Phys. \textbf{B500}, 75
(1997), \texttt{hep-th/9702103}.

\bibitem{Gibbons:1993sv}
  G.~W.~Gibbons and P.~K.~Townsend,
  \textit{Vacuum interpolation in supergravity via super p-branes},
  Phys.\ Rev.\ Lett.\  {\bf 71}, 3754 (1993), \texttt{hep-th/9307049}.

\bibitem{Sen:2005wa}
  A.~Sen,
  {\it Black hole entropy function and the attractor mechanism in higher
  derivative gravity} ,
  JHEP {\bf 0509}, 038 (2005), \texttt{hep-th/0506177}.

\bibitem{Bellucci:2007ds}
  S.~Bellucci, S.~Ferrara, R.~Kallosh and A.~Marrani,
  \textit{Extremal Black Hole and Flux Vacua Attractors}, \texttt{arXiv:0711.4547}.

\bibitem{Sen:2007qy}
  A.~Sen,
  \textit{Black Hole Entropy Function, Attractors and Precision Counting of
  Microstates}, \texttt{arXiv:0708.1270}.



\bibitem{Garousi:2007zb}
  M.~R.~Garousi and A.~Ghodsi,
  \textit{On Attractor Mechanism and Entropy Function for Non-extremal Black
  Holes/Branes,}
  JHEP {\bf 0705}, 043 (2007), \texttt{hep-th/0703260}.
M.~R.~Garousi and A.~Ghodsi,
   \textit{Entropy Function for Non-extremal D1D5 and
   D2D6NS5-branes,}
  JHEP {\bf 0710}, 036 (2007), \texttt{arXiv:0705.2149}.
  R.~G.~Cai and D.~W.~Pang,
  \textit{On Entropy Function for Supersymmetric Black Rings,
  }JHEP {\bf 0704}, 027 (2007), \texttt{hep-th/0702040}.
  K.~Goldstein and R.~P.~Jena,
 \textit{One entropy function to rule them all,
  }JHEP {\bf 0711}, 049 (2007), \texttt{hep-th/0701221}.
  K.~Hanaki, K.~Ohashi and Y.~Tachikawa,
 \textit{ Comments on charges and near-horizon data of black rings,
  }JHEP {\bf 0712}, 057 (2007, \texttt{arXiv:0704.1819}.





\bibitem{Freedman:1999gp}
  D.~Z.~Freedman, S.~S.~Gubser, K.~Pilch and N.~P.~Warner,
  \textit{Renormalization group flows from holography supersymmetry and a
  c-theorem},
  Adv.\ Theor.\ Math.\ Phys.\  {\bf 3}, 363 (1999), \texttt{hep-th/9904017}.

\bibitem{Goldstein:2005rr}
  K.~Goldstein, R.~P.~Jena, G.~Mandal and S.~P.~Trivedi,
  \textit{A $C$-function for non-supersymmetric attractors},
  JHEP {\bf 0602}, 053 (2006), \texttt{hep-th/0512138}.

\bibitem{Sen:2008vm}
  A.~Sen,
  \textit{Quantum Entropy Function from AdS(2)/CFT(1)
  Correspondence}, \texttt{arXiv:0809.3304}.


\bibitem{Brown:1986nw}
  J.D.~Brown and M.~Henneaux,
  \textit{Central Charges in the Canonical Realization of Asymptotic Symmetries: An
  Example from Three-Dimensional Gravity},
  Commun.\ Math.\ Phys.\  {\bf 104}, 207 (1986).














\bibitem{FG2}  S. Ferrara and M. G\"{u}naydin, \textit{Orbits and Attractors
for }$\mathcal{N}\mathit{=2}$\textit{\ Maxwell-Einstein Supergravity
Theories in Five Dimensions}, Nucl. Phys. \textbf{B759}, 1 (2006), \texttt{%
hep-th/0606108}.

\bibitem{Strathdee:1986jr}
  J.A.~Strathdee,
  \textit{Extended Poincar\'{e} Supersymmetry},
  Int.\ J.\ Mod.\ Phys.\  {\bf A2}, 273 (1987).

\bibitem{VanProeyen:1999ni}
  A.~Van Proeyen,
  \textit{Tools for supersymmetry}, \texttt{hep-th/9910030}.

\bibitem{Townsend:1995gp}
  P.K.~Townsend,
  \textit{P-brane democracy}, \texttt{hep-th/9507048}.





\bibitem{Ferrara-Marrani-2}  S.~Ferrara and A.~Marrani, \textit{On the Moduli
Space of non-BPS Attractors for }$\mathcal{N}\mathit{=2}$\textit{\ Symmetric
Manifolds}, Phys. Lett. \textbf{B652}, 111 (2007), \texttt{arXiv:0706.1667}.

\bibitem{AFMT1}  L. Andrianopoli, S. Ferrara, A. Marrani and M.
Trigiante, \textit{Non-BPS Attractors in 5d and 6d Extended
Supergravity}, Nucl. Phys. \textbf{B795}, 428 (2008),
\texttt{arXiv:0709.3488}.

\bibitem{Kallosh-review}  S. Bellucci, S. Ferrara, R. Kallosh and A.
Marrani, \textit{Extremal Black Hole and Flux Vacua Attractors}, to appear
in the proceedings of \textit{Winter School on Attractor Mechanism (SAM 2006)%
}, Frascati, Italy, 20-24 Mar 2006, \texttt{arXiv:0711.4547}.

\bibitem{LA08-Proc}  S. Ferrara and A. Marrani, \textit{Symmetric Spaces in
Supergravity}, contributed to \textit{Symmetry in Mathematics and Physics:
Celebrating V.S. Varadarajan's 70th Birthday}, Los Angeles, California,
18-20 Jan 2008, \texttt{arXiv:0808.3567}.



\bibitem{Tanii:1984zk}
  Y.~Tanii,
 \textit{$\mathcal{N}\mathit{=8}$ supergravity in six-dimensions},
  Phys.\ Lett.\  {\bf B145}, 197 (1984).

\bibitem{Ferrara:1997ci}
  S.~Ferrara and J.~M.~Maldacena,
  \textit{Branes, central charges and $U$-duality invariant BPS conditions},
  Class.\ Quant.\ Grav.\  {\bf 15}, 749 (1998), \texttt{hep-th/9706097}.

\bibitem{Lu:1997bg}
  H.~Lu, C.N.~Pope and K.S.~Stelle,
  \textit{Multiplet structures of BPS solitons},
  Class.\ Quant.\ Grav.\  {\bf 15}, 537 (1998), \texttt{hep-th/9708109}.

  \bibitem{ADFL6}  L. Andrianopoli, R. D'Auria, S. Ferrara and M.A. Lled\'{o},
\textit{Horizon geometry, duality and fixed scalars in
six-dimensions}, Nucl. Phys. \textbf{B528}, 218 (1998),
\texttt{hep-th/9802147}.


\bibitem{Sezgin:1982gi}
  E.~Sezgin and A.~Salam,
  \textit{Maximal extended supergravity theory in seven-dimensions},
  Phys.\ Lett.\ {\bf B118}, 359 (1982).

\bibitem{Salam:1984ft}
  A.~Salam and E.~Sezgin,
  \textit{$D = 8$ supergravity},
  Nucl.\ Phys.\ {\bf B258}, 284 (1985).



\bibitem{Andrianopoli:1997hb}
  L.~Andrianopoli, R.~D'Auria and S.~Ferrara, \textit{Five dimensional U-duality,
  black-hole entropy and topological invariants},
  Phys.\ Lett.\  {\bf B411}, 39 (1997), \texttt{hep-th/9705024}.


  \bibitem{Seiberg}  N. Seiberg, \textit{Observations on the Moduli Space of
Superconformal Field Theories}, Nucl. Phys. \textbf{B303}, 286
(1988).


\bibitem{Mo-various}  A. Belhaj, L. B. Drissi, E. H. Saidi and A. Segui, $%
\mathcal{N}\mathit{=2}$\textit{\ Supersymmetric Black Attractors in
Six and
Seven Dimensions}, Nucl. Phys. \textbf{B796}, 521 (2008), \texttt{%
arXiv:0709.0398}; E. H. Saidi, \textit{BPS and Non-BPS
}$7d$\textit{\ Black Attractors in M-Theory on }$K_{3}$,
\texttt{arXiv:0802.0583}; E. H. Saidi,
\textit{On Black Hole Effective Potential in }$\mathit{6d}$\textit{/}$%
\mathit{7d}$\textit{\ }$\mathcal{N}\mathit{=2}$\textit{\
Supergravity}, Nucl. Phys. \textbf{B803}, 235 (2008),
\texttt{arXiv:0803.0827}; E. H. Saidi and A. Segui, \textit{Entropy
of Pairs of Dual Attractors in Six and Seven Dimensions},
\texttt{arXiv:0803.2945}; A. Belhaj, \textit{On Black Objects
in Type }$\mathit{IIA}$\textit{\ Superstring Theory on Calabi-Yau Manifolds}%
, \texttt{arXiv:0809.1114}.



\bibitem{Chandrasekhar:2006kx}
  B.~Chandrasekhar, S.~Parvizi, A.~Tavanfar and H.~Yavartanoo,
  \textit{Non-supersymmetric attractors in $R^2$ gravities,
  }JHEP {\bf 0608}, 004 (2006), \texttt{hep-th/0602022}.
  A.~Sen,
  \textit{Entropy function for heterotic black holes,
 } JHEP {\bf 0603}, 008 (2006), \texttt{hep-th/0508042}.
  B.~Sahoo and A.~Sen,
 \textit{Higher derivative corrections to non-supersymmetric extremal black  holes
  in ${\cal N}=2$ supergravity,
  }JHEP {\bf 0609}, 029 (2006), \texttt{hep-th/0603149}.
  G.~L.~Cardoso, D.~L\"ust and J.~Perz,
  \textit{Entropy maximization in the presence of higher-curvature interactions,
  }JHEP {\bf 0605}, 028 (2006), \texttt{hep-th/0603211}.
  B.~Chandrasekhar,
  \textit{Born-Infeld corrections to the entropy function of heterotic black holes,
  } Braz.\ J.\ Phys.\  {\bf 37}, 349 (2007), \texttt{hep-th/0604028}.
  A.~Ghodsi,
  \textit{ $R^4$ corrections to D1D5p black hole entropy from entropy function
  formalism,
  }Phys.\ Rev.\  D {\bf 74}, 124026 (2006), \texttt{hep-th/0604106}.
  R.~G.~Cai and D.~W.~Pang,
 \textit{  Entropy function for 4-charge extremal black holes in type IIA superstring
  theory,
  }Phys.\ Rev.\  D {\bf 74}, 064031 (2006), \texttt{hep-th/0606098}.
  \textit{Entropy Function for Non-Extremal Black Holes in String Theory,
  }JHEP {\bf 0705}, 023 (2007), \texttt{hep-th/0701158}.
  D.~Astefanesei, K.~Goldstein, R.~P.~Jena, A.~Sen and S.~P.~Trivedi,
  \textit{Rotating attractors,
  }JHEP {\bf 0610}, 058 (2006), \texttt{hep-th/0606244}.
  M.~Alishahiha,
  \textit{On $R^2$ corrections for 5D black holes,
  }JHEP {\bf 0708}, 094 (2007), \texttt{hep-th/0703099}.
  A.~Castro, J.~L.~Davis, P.~Kraus and F.~Larsen,
  \textit{Precision entropy of spinning black holes,
  }JHEP {\bf 0709}, 003 (2007), \texttt{arXiv:0705.1847}.
  A.~U.~Rey,
   \textit{Contributions of Riemann invariants to the Entropy of Extremal Black
  Hole,
  }\texttt{arXiv:0811.2371}.





\bibitem{Morales:2006gm}
  J.~F.~Morales and H.~Samtleben,
  \textit{Entropy function and attractors for AdS black
  holes,}
  JHEP {\bf 0610}, 074 (2006), \texttt{hep-th/0608044}.
  X.~Gao,
  \textit{Non-supersymmetric Attractors in Born-Infeld Black Holes with a
  Cosmological Constant,}
  JHEP {\bf 0711}, 006 (2007), \texttt{arXiv:0708.1226}.
  D.~Astefanesei, H.~Nastase, H.~Yavartanoo and S.~Yun,
  \textit{Moduli flow and non-supersymmetric AdS attractors,}
  JHEP {\bf 0804}, 074 (2008), \texttt{arXiv:0711.0036}.
  P.~J.~Silva,
  \textit{On Uniqueness of supersymmetric Black holes in AdS(5),
  }Class.\ Quant.\ Grav.\  {\bf 25}, 195016 (2008), \texttt{arXiv:0712.0132}.
  J.~Choi, S.~Lee and S.~Lee,
  \textit{Near Horizon Analysis of Extremal AdS5 Black Holes,
  }JHEP {\bf 0805}, 002 (2008), \texttt{arXiv:0802.3330}.
  F.~W.~Shu and X.~H.~Ge,
  \textit{Entropy function and higher derivative corrections to entropies in
  (anti-)de Sitter space,}
  JHEP {\bf 0808}, 021 (2008), \texttt{arXiv:0804.2724}.
  D.~Astefanesei, N.~Banerjee and S.~Dutta,
  \textit{(Un)attractor black holes in higher derivative AdS gravity,
 } \texttt{arXiv:0806.1334}.




%
%
%
%


%

%
%





  \bibitem{tutti}
  R.~Kallosh,
  \textit{New attractors,
  }JHEP {\bf 0512}, 022 (2005), \texttt{hep-th/0510024}.
  P.~K.~Tripathy and S.~P.~Trivedi,
  \textit{Non-supersymmetric attractors in string theory,
  }JHEP {\bf 0603}, 022 (2006), \texttt{hep-th/0511117}.
  A.~Giryavets,
  \textit{New attractors and area codes,
  }JHEP {\bf 0603}, 020 (2006), \texttt{hep-th/0511215}.
  M.~Alishahiha and H.~Ebrahim,
  \textit{ Non-supersymmetric attractors and entropy function,
 }JHEP {\bf 0603}, 003 (2006), \texttt{hep-th/0601016}.
  R.~Kallosh, N.~Sivanandam and M.~Soroush,
  \textit{ The non-BPS black hole attractor equation,
  }JHEP {\bf 0603}, 060 (2006), \texttt{hep-th/0602005}.
  S.~Bellucci, S.~Ferrara and A.~Marrani,
  \textit{On some properties of the attractor equations,
  }Phys.\ Lett.\  {\bf B635}, 172 (2006), \texttt{hep-th/0602161}.
  S.~Ferrara and R.~Kallosh,
  \textit{On ${\cal N}=8$ attractors,
  }Phys.\ Rev.\  {\bf D73}, 125005 (2006), \texttt{hep-th/0603247}.
  S.~Bellucci, S.~Ferrara, M.~Gunaydin and A.~Marrani,
  \textit{Charge orbits of symmetric special geometries and attractors,
  }Int.\ J.\ Mod.\ Phys.\  {\bf A21}, 5043 (2006), \texttt{hep-th/0606209}.
  R.~Kallosh, N.~Sivanandam and M.~Soroush,
  \textit{Exact attractive non-BPS STU black holes,
  }Phys.\ Rev.\  {\bf D74}, 065008 (2006), \texttt{hep-th/0606263}.
  S.~Bellucci, S.~Ferrara, A.~Marrani and A.~Yeranyan,
   \textit{``Mirror Fermat Calabi-Yau Threefolds and Landau-Ginzburg Black Hole
  Attractors,''
  }Riv.\ Nuovo Cim.\  {\bf 29N5}, 1 (2006), \texttt{hep-th/0608091}.
  R.~D'Auria, S.~Ferrara and M.~Trigiante,
   \textit{Critical points of the black-hole potential for homogeneous special
  geometries,
  }JHEP {\bf 0703}, 097 (2007), \texttt{hep-th/0701090}.
  A.~Ceresole and G.~Dall'Agata,
  \textit{Flow equations for non-BPS extremal black holes,
  }JHEP {\bf 0703}, 110 (2007), \texttt{hep-th/0702088}.
  K.~Saraikin and C.~Vafa,
  \textit{Non-supersymmetric Black Holes and Topological Strings,
  }Class.\ Quant.\ Grav.\  {\bf 25}, 095007 (2008), \texttt{hep-th/0703214}.
  S.~Ferrara and A.~Marrani,
  \textit{${\cal N}=8$ non-BPS Attractors, Fixed Scalars and Magic Supergravities,
  }Nucl.\ Phys.\  {\bf B788}, 63 (2008), \texttt{arXiv:0705.3866}.
  L.~Andrianopoli, R.~D'Auria, E.~Orazi and M.~Trigiante,
  \textit{First Order Description of Black Holes in Moduli Space,
  }JHEP {\bf 0711}, 032 (2007), \texttt{arXiv:0706.0712}.
  S.~Ferrara and A.~Marrani,
  \textit{On the Moduli Space of non-BPS Attractors for ${\cal N}=2$ Symmetric Manifolds,
  }Phys.\ Lett.\  {\bf B652}, 111 (2007), \texttt{arXiv:0706.1667}.
  A.~Ceresole, S.~Ferrara and A.~Marrani,
   \textit{4d/5d Correspondence for the Black Hole Potential and its Critical
  Points,
}Class.\ Quant.\ Grav.\  {\bf 24}, 5651 (2007),
\texttt{arXiv:0707.0964}.
  K.~Hotta and T.~Kubota,
  \textit{Exact Solutions and the Attractor Mechanism in Non-BPS Black Holes,
  }Prog.\ Theor.\ Phys.\  {\bf 118}, 969 (2007), \texttt{arXiv:0707.4554}.
  D.~Gaiotto, W.~W.~Li and M.~Padi,
  \textit{Non-Supersymmetric Attractor Flow in Symmetric Spaces,
  }JHEP {\bf 0712}, 093 (2007), \texttt{arXiv:0710.1638}.
  E.~G.~Gimon, F.~Larsen and J.~Simon,
  \textit{Black Holes in Supergravity: the non-BPS Branch,
  }JHEP {\bf 0801}, 040 (2008), \texttt{arXiv:0710.4967}.
  W.~Li,
  \textit{Non-Supersymmetric Attractors in Symmetric Coset Spaces,
  }, \texttt{arXiv:0801.2536}.
  S.~Bellucci, S.~Ferrara, A.~Marrani and A.~Yeranyan,
 \textit{  d=4 Black Hole Attractors in ${\cal N}=2$ Supergravity with Fayet-Iliopoulos
  Terms,
  }Phys.\ Rev.\  {\bf D77}, 085027 (2008), \texttt{arXiv:0802.0141}.































\end{thebibliography}
\end{document}